\documentclass{aa}
\usepackage{graphicx}
\usepackage{txfonts}
\usepackage{natbib}
\bibpunct{(}{)}{;}{a}{}{,}
\begin{document}
   \title{Interpreting the galaxy group CG J1720-67.8 through evolutionary synthesis models}

   \author{S. Temporin
          \inst{1}
          \and
          U. Fritze-von Alvensleben\inst{2}
          }

   \offprints{S. Temporin}

   \institute{Institut f\"ur Astrophysik, Leopold-Franzens-Universit\"at Innsbruck, 
   Technikerstra\ss e 25, A-6020 Innsbruck, Austria\\
              \email{giovanna.temporin@uibk.ac.at}
         \and
	 Universit\"atssternwarte, Geismarlandstr. 11, D-37083 G\"ottingen, Germany\\
             \email{ufritze@uni-sw.gwdg.de}
             }

   \date{Received 5 June 2005 / Accepted 22 September 2005}

   \abstract{
   This paper is part of a series devoted to a detailed analysis of the properties of
the compact group CG J1720-67.8 and its member galaxies with the aim of sheding light
on its evolutionary history.  
Here we interpret our previously published observational results through comparison
with chemically 
consistent spectrophotometric evolutionary synthesis models in order to gain further
clues to the evolutionary history of the galaxies in this group.
In order to reduce the number of free parameters, we considered the simplest case 
of a single burst of star formation turned on after 11 - 12 Gyr of undisturbed 
galaxy evolution. However, we also briefly explore the effect of multiple, 
interaction-induced bursts of star formation.
We find that the two spiral galaxies are consistent with interaction induced strong
starbursts switched on $\sim$ 40 to 180 Myr ago and still active. For the early-type galaxy a
$\la$ 0.9 - 1.3 Gyr old star formation event (depending on the considered model) 
appears consistent with observed properties.
The comparison with models cannot rule out the possibility that this galaxy is
already the result of a merger. Alternatively, a star formation episode in this
galaxy might have been triggered by a gas inflow as a consequence of the interaction
with the companion galaxies. Estimates of galaxy masses are derived from the comparison
with the models. Finally our results are discussed in comparison with other 
well studied poor galaxy systems.

   \keywords{Galaxies: evolution --
                Galaxies: interactions --
                Galaxies: starburst
               }
   }

\authorrunning{Temporin and Fritze-v. Alvensleben}
\titlerunning{CG~J1720-67.8: Evolutionary Synthesis}
   \maketitle
%

\section{Introduction}
Hierarchical evolution theories come in different flavors depending on the assumed
cosmologies \citep[e.g. ][]{lc03,kcdw99,clbf00,spr05}, however they all suggest 
that at least part of the existing elliptical galaxies are galaxy merger products.
Such merging phenomena are thought to be particularly effective in compact galaxy groups
(CGs) today. These are in fact dense environments where the low velocity dispersion of
the galaxies --comparable to their rotational velocities-- and their small
mutual distances imply considerably short crossing times. Tidal friction can
slow down groups' members allowing them to merge, although the CGs' dark
matter content and distribution might play a fundamental role in their evolution
\citep{amb97}. Also, according to hierarchical models of structure formation,
groups are to be considered the building blocks of clusters, therefore
understanding the evolution of galaxies in the group environment is useful
to interpret the galaxy population observed in clusters.

X-ray studies have given important support to the idea that the
final fate of CGs is their coalescence into bright field elliptical galaxies,
in some of which possible CGs' relics have been identified \citep[e.g. ][]{ponm94,mz99,vik99}.
\citet{mil04} indicate X-ray dim or low velocity dispersion groups as
present sites of rapid dynamical evolution and possible modern precursors 
of fossil groups.
Nevertheless, challenges to this apparently simple scenario are not missing. In fact, recent 
studies of stellar populations of galaxies in compact groups point to
old ages of their elliptical members \citep{mdo05,pro04}, a finding at odds with the idea
that ellipticals might have formed through subsequent mergers in the dense
group environment. Obviously, the question of how groups and galaxies
within groups evolve is not settled. Investigations on groups which appear 
to be in a merging phase should help understanding these evolutionary processes.
Nearby systems in merging phase are rare to observe, but are best suited to detailed studies
that might serve as benchmarks for the interpretation of more distant galaxy systems.
This work deals with the interpretation of such a rare example of galaxy group,
caught shortly before final merging.

\object{CG J1720-67.8} is a complex galaxy system with various signs of ongoing
and past interactions. It is composed
of three main galaxies and a number of candidate tidal dwarf galaxies in extremely
dense spatial configuration and with low velocity dispersion \citep[][ hereafter Paper I]{wtk99}. 
This peculiar galaxy group is representative of the last phases of
CG evolution leading to the coalescence of the brightest members. 
The rarity of nearby systems in such a 
configuration is evidenced by the fact that the only other example of CG found 
in a similar evolutionary phase is \object{HCG 31} \citep{am04}.
In a series of papers devoted to a detailed observational study of CG J1720-67.8 
\citep[][ hereafter Paper II, III, and IV, respectively]{temp03a,temp03b,temp05},
we have presented optical photometry and spectroscopy of all group members, 
radio continuum and \ion{H}{i} observations, and far infrared emission of the group. 
We have analyzed the physical and chemical properties of the ionized gas by use of 
the observed emission-lines and CLOUDY \citep{f96} photoionization models
and investigated the ionized gas velocity field. 
In the light of these observations we have discussed
the star formation properties of the system and suggested some possible interaction 
scenarios that might have led to the present group configuration (Paper IV).
Here we concentrate on the three main galaxies and make use of chemically consistent
spectrophotometric evolutionary synthesis models to obtain the \emph{simplest} possible
interpretation of the observed properties and check their consistency with the
interaction history proposed in Paper IV. 
In particular, we aim at estimates of the ages and strengths 
of the \emph{most recent} star formation episodes, which have most likely been triggered
by galaxy interactions within the group. For our purpose we will use the lowest number 
of free parameters that provides reasonable agreement with the
available observations.
For the comparisons with models, we will make use of the observational data and derived
quantities (such as star formation rates, metallicities, extinction properties, luminosities)
that are extensively discussed in Paper~I to IV. We refer the reader to those papers for
detailed information. Here we describe only briefly the available datasets and
summarize the main properties of the galaxies (Sect. 2.1), which will be useful for the goals
of this work. We also recall the basics and input physics of the evolutionary synthesis code
{\sc galev} \citep{sch02,afva03,bfva04} that we use for our models (Sect. 2.2). In Sects. 3 and 4 
we present
and discuss the simplest models that reasonably match the observations for the late-type
and early-type galaxies, respectively. In Sect. 5 we investigate the possibility
of occurrence of multiple bursts of star formation and, finally, in Sect. 6 we discuss our results
in the framework of the possible interaction history of the group.

\section{Data and methodology}

\subsection{Observational background}

\begin{figure*}[ht]
\centering
{\large Available in jpeg format: 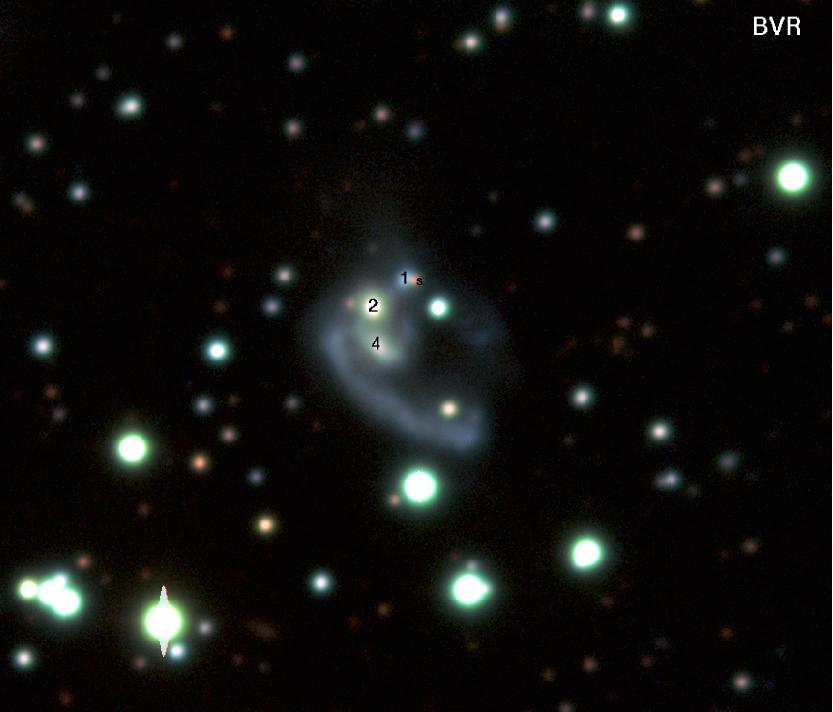}
\vskip 4in
\caption{BVR composite image of CG J1720-67.8. Numbers indicate the three main galaxies, while 
the 's' label indicate an M-type foreground star seen in projection on G1. North is on top, east
to the left. The size of the field is $\sim$ 2\farcm3$\times$1\farcm9. At the distance of the
group 1\arcsec $\sim$ 0.9 kpc. \label{RGB}}
\end{figure*}

Photometric and spectroscopic data have been obtained at the ESO 2.2 m and
3.6 m telescopes (long-slit and multi-object spectroscopy, $BVR$ imaging), in Las Campanas 
at the Du Pont 2.5 m, Swope 1 m, and Magellan 6.5 m telescopes
(long-slit spectra and $JHK_s$ imaging), and at the Anglo-Australian Observatory AAT 3.9 m 
telescope (mosaic of integral field spectra including H$\alpha$, [\ion{N}{ii}], [\ion{S}{ii}]
emission lines and adjacent continuum). Observations in the \ion{H}{i} 21 cm line and in the
radio continuum at 6, 13, and 20 cm
were obtained at the Australia Telescope Compact Array. The three main galaxies of
CG J1720-67.8, numbered 1, 2, and 4 in Papers I through IV (G1, G2, and G4, hereafter), 
are classified as an Scd, an S0,
and an Sc galaxy, respectively, on the basis of the color analysis and a quantitative
morphological study, although the strong morphological disturbances caused by their mutual
interactions make it difficult to classify these galaxies within traditional
schemes (Fig.~\ref{RGB}). 
G2 shows a typical r$^{1/4}$ bulge, while G1 and G4 have exponential bulges.
The three galaxies, as well as the candidate tidal dwarf galaxies, 
exhibit emission line spectra typical of \ion{H}{ii} or starburst galaxies.
The H$\alpha$ emission indicates considerable star formation
activity in the two spirals and a small amount of star formation essentially confined to
the central part of the S0 within a radius of $\approx$ 1.2 kpc. 
Indeed, star formation rate densities
(SFRDs) derived from the H$\alpha$ emission are comparable to and in some cases
higher than those typical of interacting galaxies (Papers I and II).
 The enhanced star formation activity
in their nuclei and throughout their disks and the clumpy distribution of star forming 
regions of G4 appear evident on the H$\alpha$ map (Paper IV). 
In fact, the high bulge to total light ratio B/T $>$ 0.4 of G1 could be a consequence
of the central star formation, which might mimic the presence of an important bulge
in an otherwise late-type-looking galaxy.
Total SFRs for the two galaxies
can be derived from their H$\alpha$ luminosities, as obtained in Paper IV (Table 5) from
integral field spectroscopic observations, after correction for internal extinction.
An average value of the extinction in the emission lines as given by the Balmer decrement 
in case B approximation is evaluated to be E($B-V$)$_{\rm g}$ $\approx$ 0.5 mag for G1 (from long-slit
spectra, see Paper II) and $\approx$ 0.6 mag for G4 (from integral field spectra, Paper IV).
Total SFRs calculated following \citet{k98} are $\sim$ 2.8 and $\sim$ 3.2 M$_{\odot}$ yr$^{-1}$ 
for G1 and G4, respectively.
We do not have information on the distribution of extinction in G1, but for G4 we found
evidence of a patchy distribution of dust and the extinction map obtained in Paper IV 
shows regions with extinction well in excess of the average value used for the correction.
Therefore, the estimated SFR of G4 is likely to be only a lower limit to the actual SFR.
This hypothesis is supported by the higher total SFR of the group  
obtained from extinction-unaffected radio continuum emission, SFR$_{\rm 1.4GHz}$ = 18 
M$_{\odot}$ yr$^{-1}$, with respect to that based on H$\alpha$ emission, 
SFR$_{\rm H\alpha}$ = 10 M$_{\odot}$ yr$^{-1}$. G4 contributes a third of the total
observed H$\alpha$ emission, but probably more than a third of the 1.4 GHz continuum emission,
since the extended radio source is peaked at G4. We therefore expect its actual
SFR to be $\ga$ 6 M$_{\odot}$ yr$^{-1}$. The above arguments must be considered when 
comparing the observed SFR with values given by the evolutionary synthesis models.
The integrated H$\alpha$ emission-line flux of G2 (see Paper~IV, Tables 1 and 5), 
when corrected for internal extinction, implies a
SFR\footnote{uncorrected values would be 0.1 to 0.2 M$_{\odot}$ yr$^{-1}$.} 
$\sim$ 0.4 to 0.8 M$_{\odot}$ yr$^{-1}$.
The three galaxies appear in mutual interaction and a bright, knotty tidal tail
departing from G4 extends for $\approx$ 29 kpc (H$_0$ = 75 km s$^{-1}$ Mpc$^{-1}$).
From the maximum amplitude of the velocity curve of G4 and the projected length of
the tidal tail we have estimated for it an age of $\approx$ 200 Myr (Paper IV). 
All the group's members show blue colors, with the exception of G2 whose
observed red color ($B-V$ = 0.98) is probably a combined effect of the high internal reddening
(A$_V$ = 1.6 mag, as derived from the Balmer decrement in its spectrum) and of its metallicity 
(Z $\sim$ 0.5 Z$_{\sun}$, based on emission line measurements), higher than that
of all other group's members, which have Z $\sim$ 0.3 Z$_{\sun}$ or lower (see Papers
II and III).

\subsection{Evolutionary synthesis with the code {\sc GALEV}}

We use the G\"ottingen evolutionary synthesis code {\sc galev} as first
presented by \citet{fvag94} and improved/updated by \citet{ku99} and then by
\citet{sch02}. The version of the code we applied is based on the Padova 
isochrones for metallicities\footnote{Z is the abundance by mass of
elements more massive than $^4$He; Z$_{\sun}$ = 0.018} in the range 10$^{-4}$ $\leq$ 
Z $\leq$ 0.05 
\citep[ as obtained from the Padova web server in the extended and improved version
of November 1999]{be94}, extended toward
lower stellar masses  (0.15 - 0.45 M$_{\sun}$) by Zero Age Main Sequence points
from \citet{cb97}.
A complete library of model atmosphere spectra --in terms of spectral types from 
hottest to coolest stars and all luminosity classes-- from \citet{lcb97,lcb98}
is implemented for the full range of metallicities spanned by the isochrones.
Spectra of all states along an isochrone for a given metallicity are summed up
(since this is a one-zone model without any spatial resolution) and weighted with the initial
mass function (IMF). Two different IMFs are available, a \citet{salp55} IMF and a
\citet{sc86} IMF, the latter parametrized as $\phi$(m) $\sim$ m$^{-x}$ with $x$ =
$-$1.25 for m$_l$ $\leq$ m $\leq$ 1 M$_{\sun}$, $x$ = $-$2.35 for 1 M$_{\sun}$ $\leq$
m $\leq$ 2 M$_{\sun}$, and $x$ = $-$3.00 for 2 M$_{\sun}$ $\leq$ m $\leq$ m$_u$.
For both IMFs the lower and upper mass limits m$_l$ and m$_u$ are 0.08 M$_{\sun}$
and 85 M$_{\sun}$, respectively.
In our models we adopt the Scalo IMF and assume a fraction of visible mass
FVM = 0.5, which gives mass-to-light ratios for undisturbed galaxies after $\sim$ 12 Gyr 
in agreement with observations
and implies that 50\% of the mass transformed into stars is in the form of Jupiter-like
objects or brown dwarfs \citep{fvag94}.
Nebular emission is not included in the version of the code we used, however the
influence of nebular emission on colors has been evaluated by \citet{afva03} 
using an updated version of {\sc galev}. In particular, the impact of gaseous
emission on broad band colors is found to be negligible, except for very
strong bursts in very early stages (ages $<$ few 10$^6$ yr up to $\sim$ 2$\times$ 10$^7$ yr, 
slightly depending on metallicity).
Although {\sc galev} allows for evolutionary and cosmological
corrections, including attenuation effects by intergalactic hydrogen, we did not
apply them, since they are unimportant at the relatively low redshift (z =
0.045) of CG~J1720-67.8.

The model calculations start with a proto-galaxy of given total mass, maintained
constant during the whole evolution (closed-box model). The metallicity Z of the 
stellar population
can be chosen between five different values (0.0004, 0.004, 0.008, 0.02, and
0.05) or calculated in a chemically consistent (CC) way, i.e. accounting for
the increasing initial metallicity of successive generations of stars \citep{fva99}. 
%

\begin{figure}[ht]
\centering
\includegraphics[width=\linewidth]{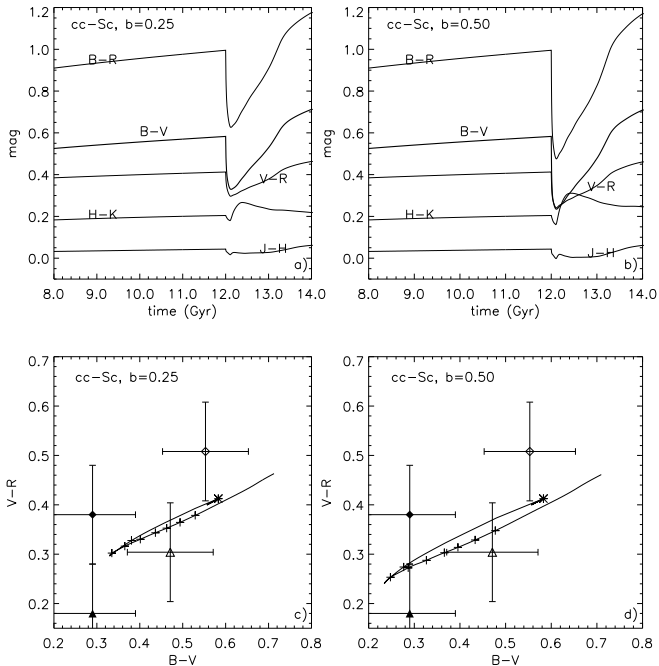}       
\caption{{\bf a)} Evolution of color indices for a cc-Sc model with a 25\% burst 
switched on after 12 Gyr of undisturbed evolution. The $B-V$ index is indicated with a thick line to avoid
confusion with other indices. For the same reason, an offset of $-$0.5 has been applied to
the $J-H$ index. {\bf b)} Same as {\bf a)}, for a 50\% burst.
{\bf c)} Color-color diagram showing the color indices evolution during the burst phase for
a cc-Sc model with a 25\% burst. The asterisk marks the onset of the burst at a galaxy age of
12 Gyr. Plus symbols are plotted for burst ages from 40 Myr to 1 Gyr at intervals of 140 Myr.
The empty triangle and diamond indicate the observed colors of G1 and G4, respectively, after a standard
inclination correction. The full triangle and diamond indicate the observed colors after correction for
internal extinction using E($B-V$) = 0.5$\times$E($B-V$)$_{\rm g}$, following \citet{cks94}.
{\bf d)} Same as {\bf c)} for a 50\% burst model. \label{colevolSc}}
\end{figure}
In the presence of starburst-driven galactic scale winds or gas accretion events, 
the closed-box representation might be inadequate and in such a situation 
it could be useful to assume a fixed metallicity (i.e. non-CC model) rather than to
introduce an exceedingly high number of additional free parameters in a CC model
(e.g. in/outflow rate, timescale, metallicity, and time evolution).
In both non-CC and CC models the
initial abundance in the gas is set to Z(0) = 10$^{-4}$. Chemical composition
gradients in the galaxy are neglected. The increase
of gas metallicity in time and the gas content are calculated by solving a
modified form of Tinsley's equations \citep{tin80} with stellar yields for type I
and II supernovae, planetary nebulae, and stellar mass loss \citep{li99}. 
The composite stellar population of a galaxy is obtained by folding models for
single stellar populations with a particular star formation history (SFH).
Different SFHs are adopted appropriately for different spectral galaxy types.  
An exponential parametrization of the star formation rate (SFR), 
$\Psi(t)$ $\sim$ $\exp(-t/t_{\ast})$ is used for elliptical galaxies, a linear 
function of the gas-to-total mass ratio, $\Psi(t)$ $\sim$ $\frac{\rm G}{\rm M}(t)$, 
is assumed for Sa to Sc spirals \citep[according to what was found by e.g. ][]{ken98}.
The constant of proportionality (i.e. the efficiency factor) is chosen to yield 
characteristic gas consumption timescales for the respective spiral types and 
decreases going from S0 to late-type spirals. Except for the fact that some gas
is given back as soon as the first stars die, this is similar
to the timescale for the decrease of the SFR from its initial value. This initial SFR is
highest for ellipticals and becomes progressively lower for later types.
A constant SFR is assumed for Sd galaxies.
 Characteristic time scales for star formation are 1 Gyr for ellipticals and 2, 3, 
 10, and 16 Gyr for Sa, Sb, Sc and Sd galaxies, respectively. Several quantities 
 are calculated on the basis of these SFHs
 at every timestep of the galaxy evolution: luminosities, color indices, star
 and gas masses, metallicities, ejection rates of gas and metals by dying stars,
 SFR, and the spectra in the wavelength range from UV to NIR. 
 The SFHs are chosen such as to bring the
   colors, spectra, and gas content of the model galaxies after 12 Gyr of undisturbed evolution
   into agreement with the average observed colors, template spectra, and 
   typical gas fraction of the respective
   galaxy types in the RC3 catalog \citep{bu95,bw95,k92}.
 Galaxy luminosities in $B$ are normalized to the average observed $B$-band luminosity 
 $\langle$M$_{B}\rangle$ of the respective galaxy types in Virgo cluster \citep{sa85}
 after 12 Gyr of evolution. More detailed explanations as well as galaxy model spectra
 for various spectral types and ages in the range 4 Myr to 15 Gyr are presented by \citet{bfva04}.
 On top of the
 normal evolution, one or more bursts of star formation can be implemented. A
 starburst is simulated by adding an exponential SFR with an e-folding timescale $t_{\ast}$
 of order 10$^8$ yr and with a maximum burst strength governed by the gas
 reservoir of the galaxy at the time of the onset of the burst. The burst strength is
 quantified by the parameter $b$ defined as the ratio of the total stellar mass
 formed during the burst (i.e. integrated until the time at which the burst is exhausted,
 the residual SFR is too low to balance the return of mass to the gas reservoir through
 stellar winds, and the total stellar mass does not increase any further) to the mass of stars 
 present in the galaxy at the onset of the burst.

\begin{figure*}[ht]
\centering
\hbox{
\includegraphics[width=0.5\linewidth]{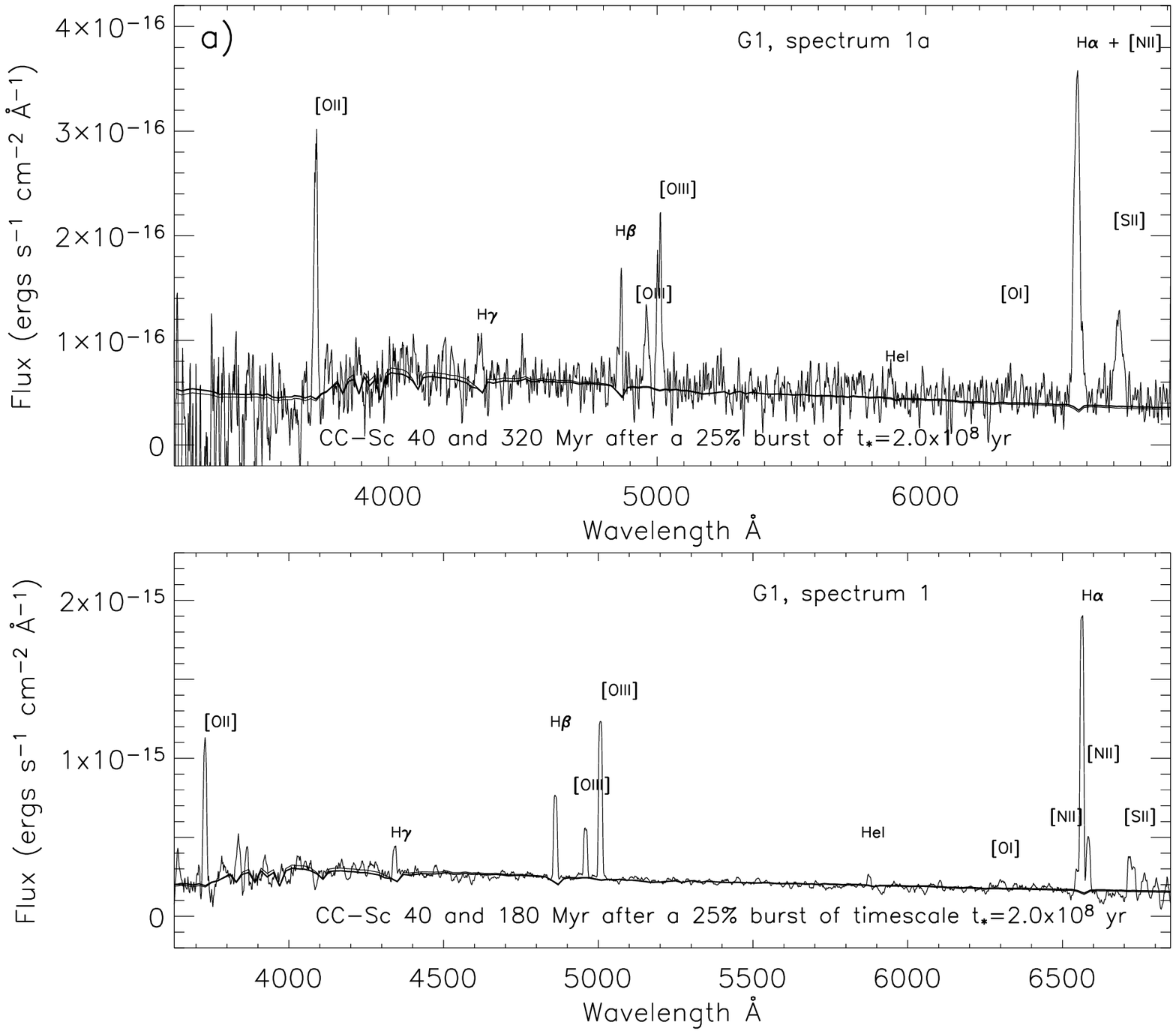}    
\includegraphics[width=0.5\linewidth]{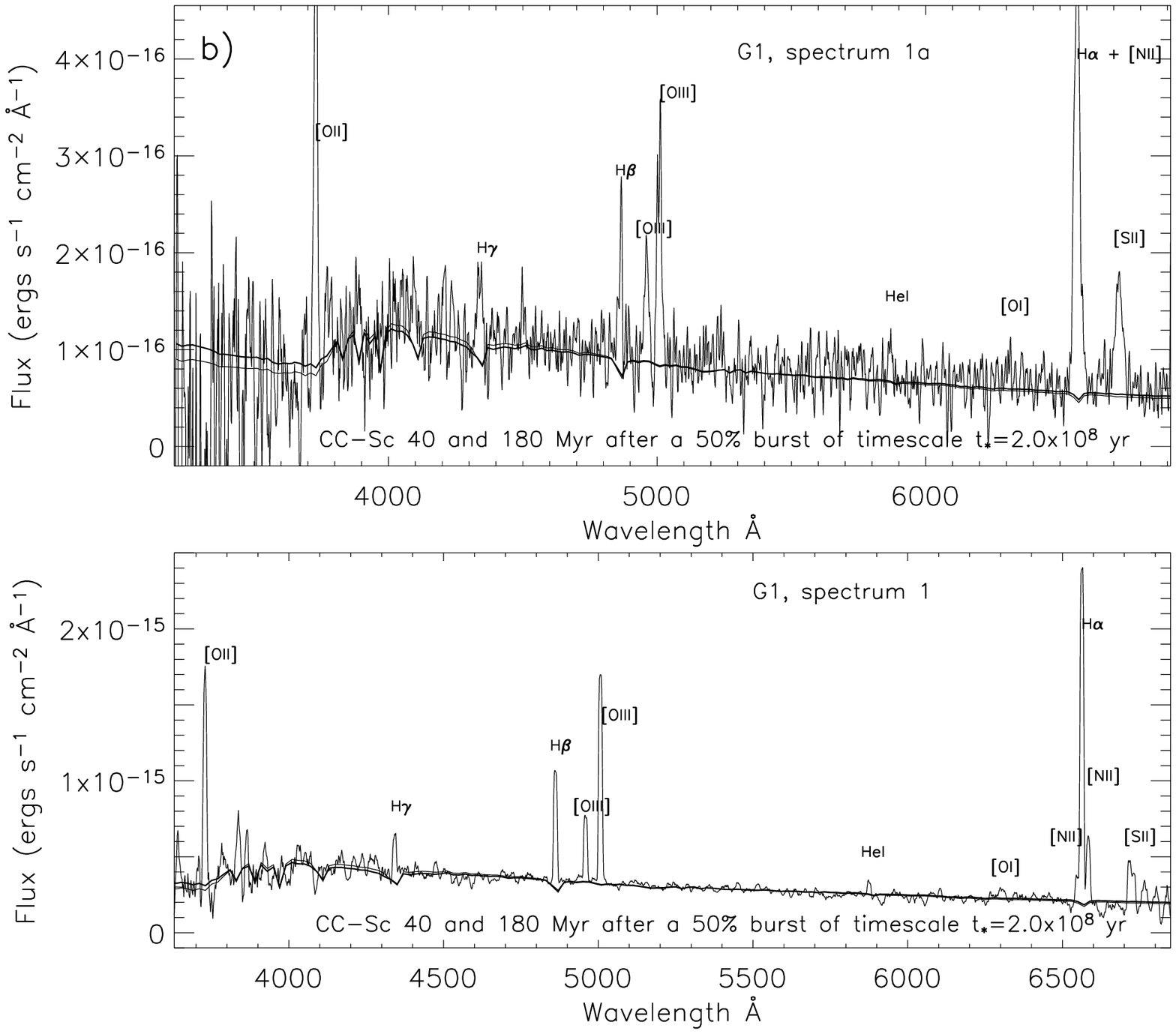}}  
\caption{{\bf a)} CC-Sc model spectra calculated 40 (thick line) and 180 or 320 (thin lines) Myr after a  
25\% burst compared with the two spectra of G1 at P.A. 90\degr\ (\emph{top}) and 130\degr (\emph{bottom}) after 
internal reddening correction (see text). Displayed models are sufficiently good approximation of the observed
spectra, with reduced $\chi^2$ in the range 0.12 - 0.16, however best fits were obtained for exceedingly low values
of internal extinction with respect to initial estimates based on Balmer decrement. 
Spectrum 1, as presented in Paper II, was contaminated by an M-type star overlapped 
to G1 in projection on the sky. 
For the purpose of comparison with models we have re-extracted the one-dimensional spectrum
excluding the contaminated portion. For this reason, the total intensity of the spectrum
and the continuum shape (in the red) differ from those shown in Paper II.
{\bf b)} Same as {\bf a)} for a model with a 50\% burst. Model spectra at both burst ages
acceptably reproduce the observed spectra. A minimum reduced $\chi^2$ = 0.12 is obtained for both
spectra at age 180 Myr, while the 40 Myr model gives a reduced $\chi^2$ = 0.16 and 0.15 
for spectra 1a and 1, respectively. Best fit extinction values are E($B-V$) = 0.30 - 0.35 and
0.12 mag for spectra 1a and 1, respectively. \label{g1CC1spec}}
\end{figure*}

\section{Late-type group members: G1 and G4}

Our analysis in Paper II showed that G1 and G4
have photometric and spectroscopic characteristics consistent with Scd and Sc 
morphological types, respectively, although none of these galaxies shows
clear spiral arms and G1 has a high bulge-to-total light ratio.
We interpret these morphological characteristics as an indication of an 
undergoing transition of the galaxies toward earlier morphological types
as an effect of their interactions. Bulge-to-disk ratios and spectral types
can be affected by strong bursts, therefore caution is necesary when choosing
the SFH to be adopted in the models. Given the above considerations, we
judged it reasonable to assume SFHs typical of late-type spirals, i.e. Sd or Sc,
when modeling G1 and G4. Since modeling attempts with an Sd-type SFH did not
produce satisfactory results\footnote{Models which assume an Sd-type SFH 
produce similar results to those presented in this Section for the Sc-type SFH
concerning the burst strengths, the best-fit burst ages, the best-fit
internal extinctions, and the mass estimates for G1 and G4,
however they predict too high SFRs and metallicities with respect to the observations and, at the best-fit
ages, they show disagreements with the central part of the observed optical spectra and 
with the observed optical-NIR spectral energy distribution. Disagreements are more significant 
for G4 than for G1, as expected from their morphological classification.}, 
we present here only models that assume an Sc-type SFH.

In our models we simulated undisturbed evolution
of an Sc galaxy for 12 Gyr (which accounts for the older stellar populations of the
galaxies), then added a starburst event on top of the normal evolution.
The color evolution of the model after the onset of the burst, calculated with a timestep
of 4 Myr, was compared with optical/NIR 
observed colors of the galaxies to obtain a first rough indication of the burst strength and age.
The timescale of the starburst should be of the order of a dynamical timescale in
the burst region. For a given burst strength, changing the timescale of the burst would not
change remarkably the shape of the optical spectrum, therefore would have little influence
on the determination of the burst age, while it would affect the luminosity at a given age
thus influencing the final estimate of the total mass of the galaxy. In our models we
assumed a burst timescale of 2$\times$10$^8$ yr. Models with timescale 1$\times$10$^8$ yr
would lead to the same burst ages estimated below and, for the ages of interest, would give 
differences in absolute magnitudes in the range 0.01 - 0.05 mag and would imply lower 
total masses of the galaxies by a factor $\approx$ 1.2.

\begin{figure}
\centering
\includegraphics[width=\columnwidth]{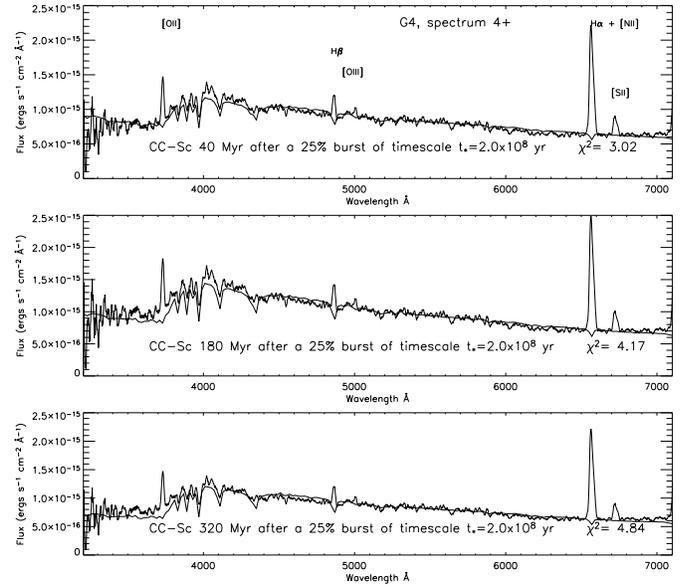} 
\caption{CC-Sc model spectra calculated 40,180, and 320 Myr (from top to bottom, thick line) after a  
25\% burst compared with the spectrum of G4 (at P.A. = 90\degr) after 
internal reddening correction (see text). The 40 Myr model is the one better approaching the observations,
taking into account an internal extinction E($B-V$) = 0.39 mag. 
The reduced $\chi^2$ of the fit based on continuum intervals and absorption line equivalent withs,
minimized over a range of possible internal extinction values, is indicated for the 3 model ages.
Best fits were obtained correcting for an internal extinction in the range E($B-V$) = 0.39 (younger ages) - 
0.44 mag (older age).
\label{g4+CC1spec}}
\end{figure}

\begin{figure*}
\centering
\hbox{
\includegraphics[width=0.5\linewidth]{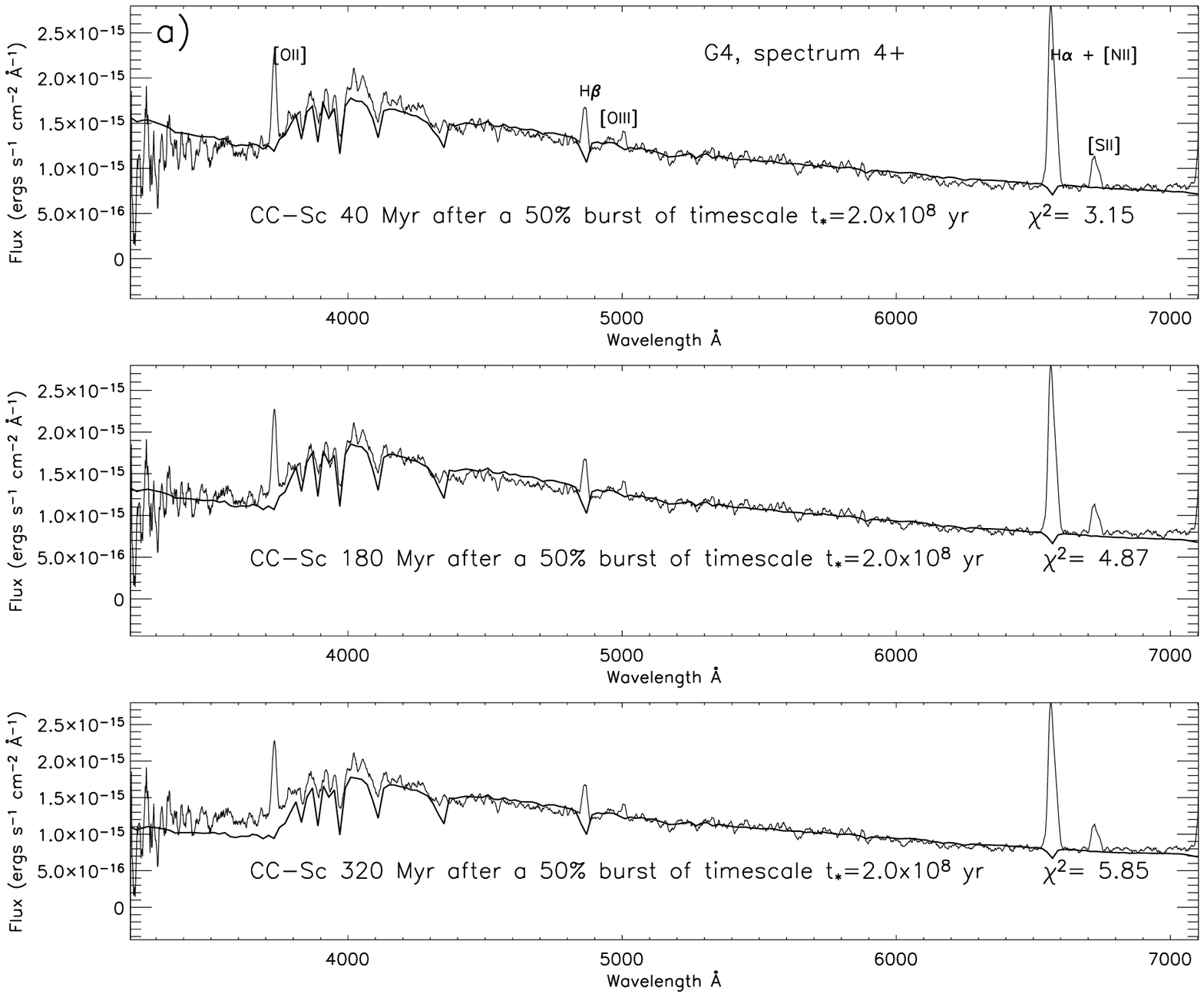}   
\includegraphics[width=0.5\linewidth]{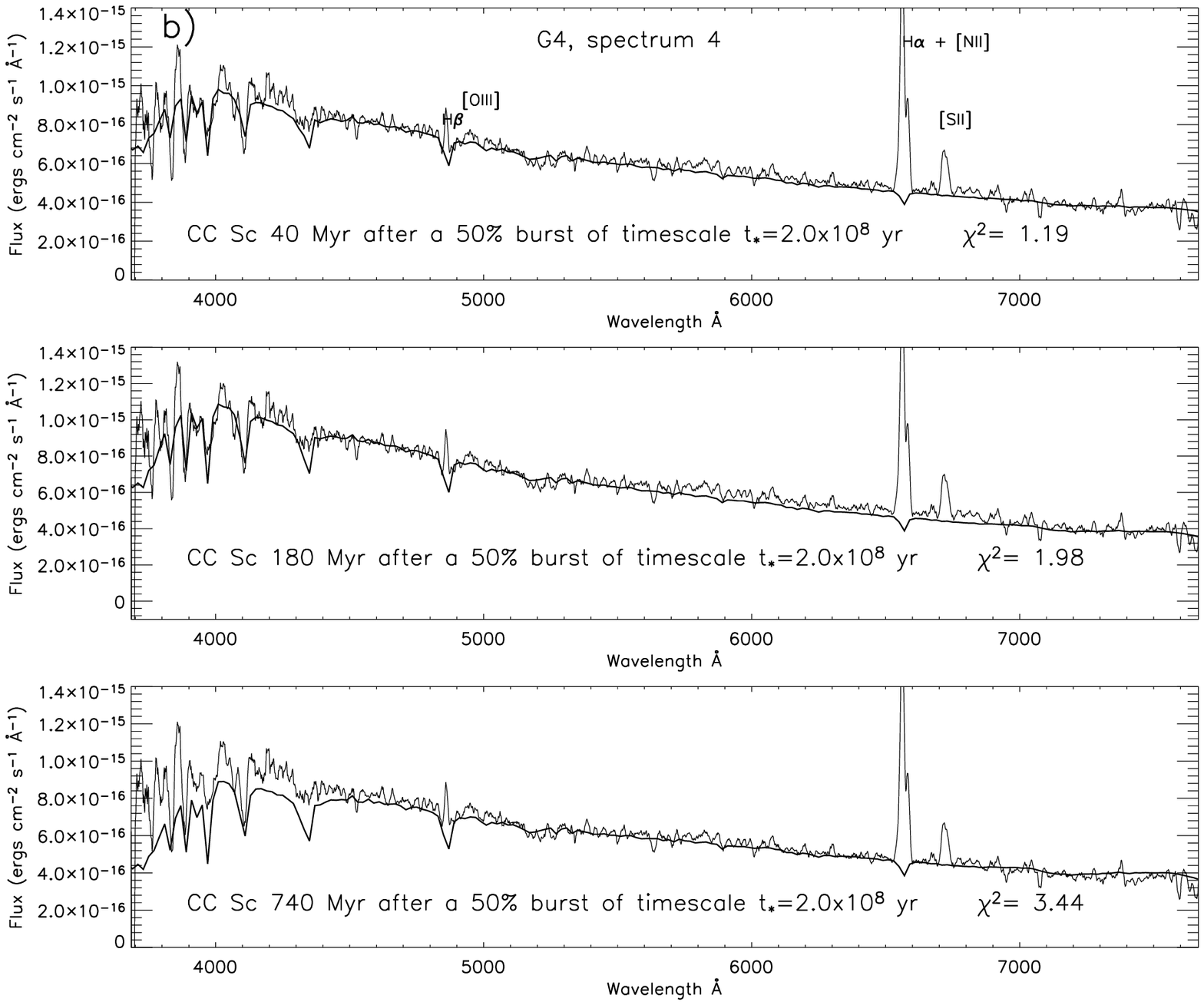}}  
\caption{{\bf a)} CC-Sc model spectra calculated 40,180, and 320 Myr (from top to bottom, thick line) after a  
50\% burst compared with the spectrum of G4 at P.A. = 90\degr, after 
internal reddening correction (see text). Best fits were obtained considering an internal extinction
E($B-V$) = 0.49 mag. Reduced $\chi^2$ values are indicated. The first two models are in good agreement 
with the data, while the third one departs from the observations in the blue part of the spectrum; 
the minimum 
reduced $\chi^2$ is obtained for the 40 Myr burst age.
{\bf b)} Same model as in {\bf a)} compared to the spectrum 
of G4 at P.A. = 130\degr\ for burst ages 40, 180, and 740 Myr. 
The internal extinction E($B-V$) determined through the fitting procedure is in the range 0.33 - 0.35 mag. 
Reduced-$\chi^2$ values are marked. The older burst model departs significantly from
the observations while the first two models match the data reasonably well; the minimum 
reduced $\chi^2$ is obtained for the 40 Myr burst age.\label{g4+CC2spec}}
\end{figure*}

 Out of several burst 
strengths we tried, we show here the results for a moderate ($b$ = 25\%) and a strong 
($b$ = 50\%, the maximum strength achievable without exhausting the gas reservoir) burst model. 
These models represent the simplest case, in which the 
presently observed burst of star formation is the first interaction-triggered star 
formation event in these galaxies. We will discuss in Sect. 4 the possibility of occurrence
of multiple starbursts.
The evolution of a few color indices is shown in Fig.~\ref{colevolSc} for
the two burst strengths considered here. Color indices during the burst phase are to be
compared with observed galaxy colors given in Table~\ref{colorsg1g4}. We also show
an example of a color index loop in the color-color diagram during the burst phase.
The beginning of the burst and subsequent 140 Myr timesteps have been marked
along the color loop for burst ages in the range 40 Myr - 1 Gyr.
For comparison, observed colors of G1 and G4 are shown after the application of a 
standard inclination correction (empty symbols; these correction are reported in Paper II
and are supposed to account for the internal extinction expected for a given morphological
type), and after a correction
for internal extinction as determined from observed spectra, following \citet{cks94} as 
explained below (full symbols). When the effects of internal extinction are taken into account,
the strong burst model appears more suitable to explain the observations.
Considering also other color indices we found that observed colors appear consistent
with a burst age in the range $\sim$ 8 -- 320 Myr. However, errors in the color determination
are quite large, as visible from the error bars in Fig.~\ref{colevolSc}, and corrections
for internal extinction of total galaxy magnitudes/colors are highly uncertain.
The use of spectral features is particularly useful to improve this initial rough estimate
of the burst age. Therefore, 
synthetic spectra calculated at intervals of 140 Myr were compared with observed spectra
within the age interval of interest. Before comparison, a correction for internal extinction
had to be applied to the observed spectra. As shown by \citet{cks94} and \citet{calz01}, the UV and optical
continuum of starburst galaxies suffers from lower extinction than the emission lines, 
in particular the optical depth of the continuum underlying the Balmer lines is found to be 
about one-half of the difference between the optical depths of the Balmer emission lines.
Therefore, as an initial estimate of the reddening affecting the optical continuum 
we assumed E($B-V$) = 0.5$\times$E($B-V$)$_{\rm g}$. We then varied E($B-V$) at steps of 0.05 mag
(or smaller) around this value. We applied this range of possible corrections by means of
the {\sc iraf} task {\sc deredden} assuming a \citet{ccm89} extinction law with selective
extinction R$_V$ = 3.1. The resulting set of corrected spectra, normalized to their continuum
flux around 5000 \AA, was compared with normalized model spectra by means of an automated procedure
based on $\chi^2$ evaluation. The average flux within several (10 to 12) intervals of continuum and 
the equivalent widths (EW) of 15 absorption lines\footnote{In G1's spectra, absorption lines are
very weak and in most cases hidden by emission lines, so only continuum intervals were considered for this galaxy.} 
were used for the calculation of the reduced-$\chi^2$.
For each age in the interval of interest, the relevant model spectrum was compared with the whole set
of observed spectra corrected for differing values of internal extinction and the one giving the
minimum reduced-$\chi^2$ was chosen. This way, each age had an associated $\chi^2$ and optimal value
of E($B-V$). The age with minimum reduced-$\chi^2$ value was chosen as best representation of the
observed spectrum. In some cases the variation of reduced-$\chi^2$ between models at adjacent burst age values
can be very small. In these circumstances, any of these ages could be reliable, not only the one
with minimum reduced-$\chi^2$. 
The final best values of E($B-V$) in case of $b$ = 50\% models were found to be
0.12 mag and 0.30 - 0.35 mag for spectra 1 (P.A. = 130\degr) and 1a (P.A. = 90\degr, sampling a different
portion of the galaxy, see Paper II) of G1, respectively. For the two available spectra of G4 (spectra 4 and 4+ at 
P.A. = 130\degr\ and 90\degr) we found E($B-V$) = 0.35 and 0.49 mag, respectively, 
in agreement with the expectations from the measured E($B-V$)$_{\rm g}$, taking into account the errors.
Comparably good fits could be obtained with lower E($B-V$) values and $b$ = 25\% models.
While for G4 these lower extinction values, E($B-V$) = 0.23 and 0.39 mag for the two spectra, could
still be acceptable within errors, in the case of G1, especially for spectrum 1, the best
fit to the moderate burst model would require far too low an extinction value,
about 1/6 of the measured E($B-V$)$_{\rm g}$. This is a further indication that the moderate burst
is not an adequate model for G1.
In Figs.~\ref{g1CC1spec} to \ref{g4+CC2spec}, we show the comparison between 
the extinction corrected
spectra of G1 and G4 and model spectra, conveniently scaled in intensity, for the two 
assumed burst strengths. Since the spectral resolution of the models (20 \AA) is lower than that of
observed spectra, a box-car smoothing was applied to the latter.

In collapsing along the spatial direction spectrum 1 of G1, we have 
excluded a few columns contaminated by the overlapping spectrum of an M-type star, which would
compromise the comparison with model spectra. In case of both moderate and strong bursts
(Figs.~\ref{g1CC1spec}{\bf a)} and {\bf b)}), the
models which better approach the observed spectra of G1 are those at a burst age in the range 40 - 180 
Myr, with the oldest age favored by the calculated reduced-$\chi^2$.

\begin{table*}
\footnotesize
\caption[]{G1 and G4: Observed$^{\mathrm{a}}$ and Modeled Colors}
\label{colorsg1g4}
$$
\begin{tabular}{llllllllll}

\hline
\hline
\noalign{\smallskip}
Object/Model & $B-V$ & $V-R$ & $V-K$ & $B-R$ & $B-H$ & $J-H$ & $H-K$ & Z/Z$_{\sun}$ \\
\noalign{\smallskip}
\hline
\noalign{\smallskip}
G1 & 0.47 & 0.30 & 2.43 & 0.78 & 2.81 & 0.43 & 0.09 & $\sim$0.2\\
G4 & 0.55 & 0.51 & 2.20 & 1.06 & 2.57 & 0.48 & 0.18 & $\sim$0.1\\
 & & & & & & & & \\
CC-Sc, $b$=0.5, 40 Myr  & 0.26 & 0.35 & 1.91 & 0.61 & 1.98 & 0.49 & 0.19 & 0.22\\
CC-Sc, $b$=0.5, 180 Myr&0.23 & 0.31 & 1.97 & 0.54 & 1.99 & 0.53 & 0.21 & 0.30\\
CC-Sc, $b$=0.5, 320 Myr&0.27 & 0.34 & 2.11 & 0.60 & 2.15 & 0.55 & 0.23 & 0.37\\
 & & & & & & & & \\
CC-Sc, $b$=0.25, 40 Myr& 0.36 & 0.41 & 2.08 & 0.77 & 2.25 & 0.51 & 0.20 & 0.21\\
CC-Sc, $b$=0.25, 180 Myr& 0.32 & 0.37 & 2.08 & 0.69 & 2.19 & 0.53 & 0.21  & 0.24\\
CC-Sc, $b$=0.25, 320 Myr& 0.35 & 0.39 & 2.18 & 0.74 & 2.31 & 0.55 & 0.22  & 0.26\\
\noalign{\smallskip}
\hline
\end{tabular}
$$
\begin{list}{}{}
\item[$^{\mathrm{a}}$] Observed colors after the application of a standard inclination correction.
\end{list}
\end{table*}

\begin{table*}
\caption[]{G1 and G4: Observed$^{\mathrm{a}}$ and Modeled Luminosities}
\label{lumg1g4}
$$
\begin{tabular}{llllllll}
\hline
\hline
\noalign{\smallskip}
Object/Model & M$_{B}$ &  M$_{V}$ & M$_{R}$ & M$_{J}$ & M$_{H}$ & M$_{K}$ & SFR (M$_{\odot}$ yr$^{-1}$)\\
\noalign{\smallskip}
\hline
\noalign{\smallskip}
G1 & $-$18.92 & $-$19.39 & $-$19.70 & $-$21.30 & $-$21.73 & $-$21.82 & 2.8\\
 & & & & & & & \\
CC-Sc, $b$=0.5 & & & & & & & \\
 40 Myr, M = 5.96$\times$10$^{9}$ M$_{\odot}$ & $-$19.09 & $-$19.35 & $-$19.70 & $-$20.58 & $-$21.07 & $-$21.26 & 4.3 \\
180 Myr, M = 4.12$\times$10$^{9}$ M$_{\odot}$ & $-$19.15 & $-$19.38 & $-$19.70 & $-$20.62 & $-$21.14 & $-$21.35 & 1.5\\
320 Myr, M = 3.92$\times$10$^{9}$ M$_{\odot}$ & $-$19.10 & $-$19.37 & $-$19.70 & $-$20.69 & $-$21.25 & $-$21.48 & 0.69\\
 & & & & & & & \\
CC-Sc, $b$=0.25 & & & & & & \\
 40 Myr, M = 7.23$\times$10$^{9}$ M$_{\odot}$ & $-$18.92 & $-$19.29 & $-$19.70 & $-$20.66 & $-$21.17 & $-$21.37 & 2.6\\
180 Myr, M = 5.74$\times$10$^{9}$ M$_{\odot}$ & $-$19.01 & $-$19.32 & $-$19.70 & $-$20.67 & $-$21.20 & $-$21.40 & 1.0\\
320 Myr, M = 5.57$\times$10$^{9}$ M$_{\odot}$ & $-$18.97 & $-$19.31 & $-$19.70 & $-$20.72 & $-$21.27 & $-$21.49 & 0.49\\
 & & & & & & & \\
 G4 & $-$20.30 & $-$20.86 & $-$21.36 & $-$22.40 & $-$22.87 & $-$23.06 & 3.2 \\
  & & & & & & & \\
CC-Sc, $b$=0.5& & & & & & & \\  
40 Myr,  M = 3.11$\times$10$^{10}$ M$_{\odot}$ & $-$20.88 & $-$21.15 & $-$21.49 & $-$22.38 & $-$22.87 & $-$23.06 & 22.1\\
180 Myr, M = 1.98$\times$10$^{10}$ M$_{\odot}$ & $-$20.88 & $-$21.09 & $-$21.40 & $-$22.32 & $-$22.85 & $-$23.06 & 7.0\\
320 Myr, M = 1.68$\times$10$^{10}$ M$_{\odot}$ & $-$20.68 & $-$20.94 & $-$21.28 & $-$22.27 & $-$22.83 & $-$23.06 & 2.9\\
 & & & & & & & \\
 CC-Sc, $b$=0.25& & & & & & & \\  
 40 Myr,  M = 3.44$\times$10$^{10}$ M$_{\odot}$ & $-$20.61 & $-$20.98 & $-$21.39 & $-$22.35 & $-$22.86 & $-$23.06 & 12.2\\
 180 Myr, M = 2.64$\times$10$^{10}$ M$_{\odot}$ &$-$20.66 & $-$20.97 & $-$21.35 & $-$22.32 & $-$22.85 & $-$23.06 & 4.6\\
 320 Myr, M = 2.38$\times$10$^{10}$ M$_{\odot}$ &$-$20.54 & $-$20.89 & $-$21.27 & $-$22.30 & $-$22.85 & $-$23.06 & 2.1\\
\noalign{\smallskip}
\hline
\end{tabular}
$$
\begin{list}{}{}
\item[$^{\mathrm{a}}$] Observed luminosities after the application of a standard inclination correction.
\end{list}
\end{table*}

For G4, in Figs.~\ref{g4+CC1spec} and \ref{g4+CC2spec}, we show the comparison with spectra nos. 4 
and 4+. In these spectra, the most important
absorption lines are detected and the \ion{Ca}{ii} absorption index \citep{ro84,lr96} is typical 
of A0- or B-type stars.
Also in this case, model spectra calculated 40 and 180 Myr after the strong burst are those better 
approaching the observed spectra, the youngest age being favored by the reduced-$\chi^2$. 
The model spectra at burst ages of 320 and 740 Myr are also displayed to show how 
models with older burst ages ($\geq$ 320 Myr) depart from the observations. 

\begin{figure*}
\centering
\hbox{
\includegraphics[width=0.5\linewidth,height=8cm]{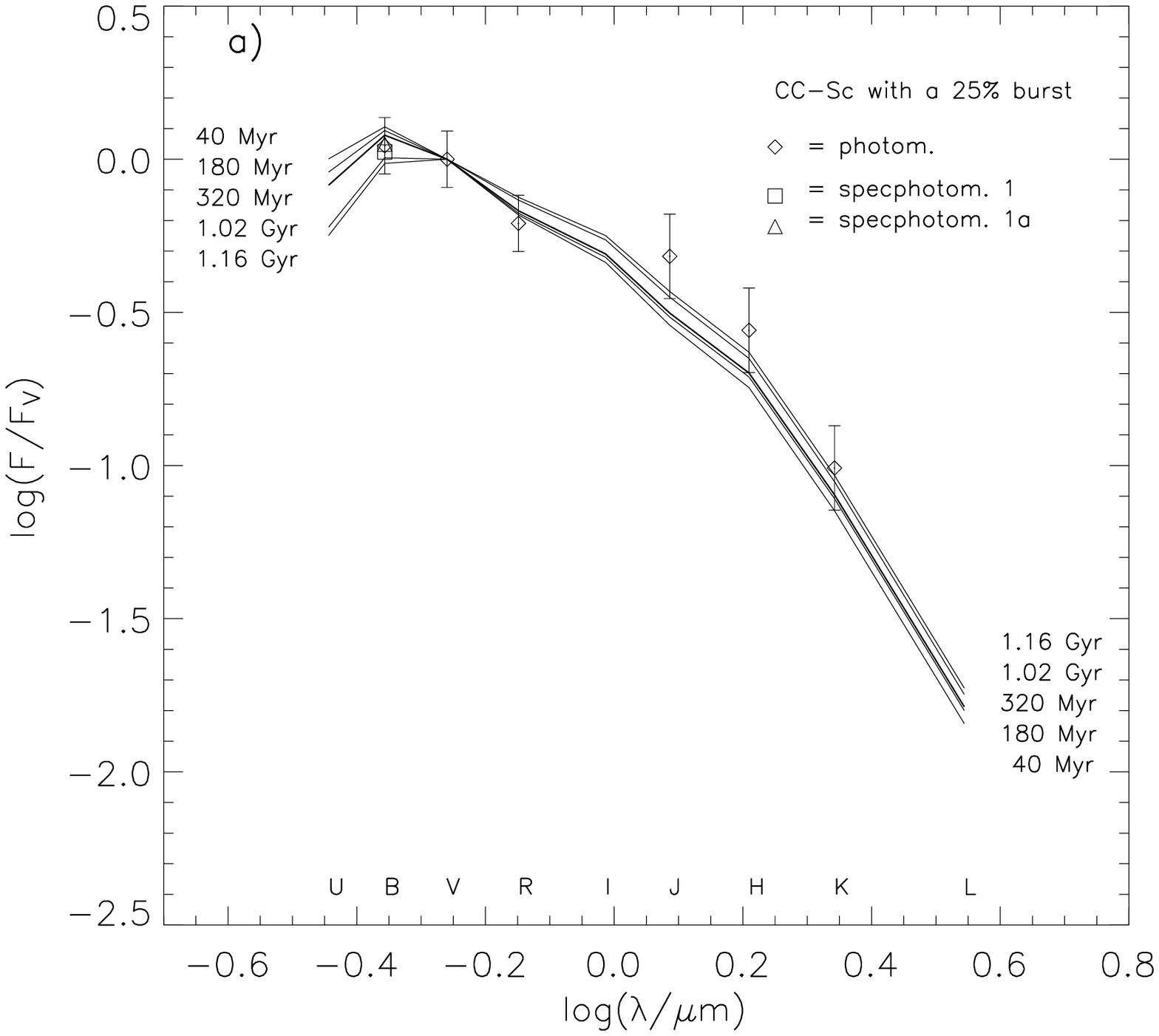}  
\includegraphics[width=0.5\linewidth,height=8cm]{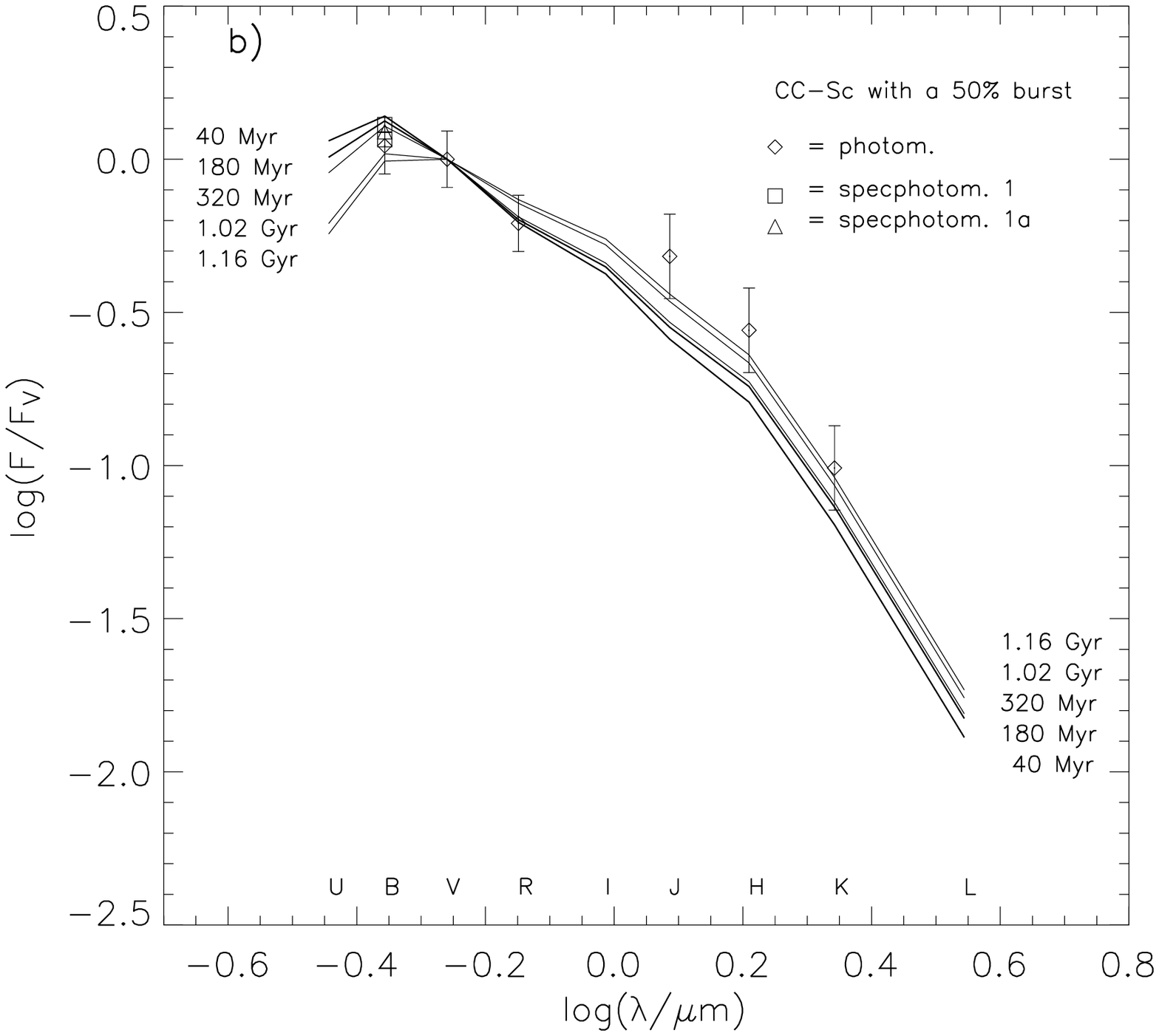}} 
\caption{Energy distribution normalized to the $V$ band for CC-Sc models in a wide range
of burst ages after the onset of a 25\% 
[{\bf a)}] and a 50\% [{\bf b)}] burst. Diamonds with associated
error bars are photometric observational points for G1, after
correction for inclination. The square and the triangle are
spectrophotometric points obtained from the  observed, dereddened,
spectra nos. 1 and 1a after convolution with $B$ and $V$ filter response
functions \citep{bess90}. The discrepancy with NIR data points might be
an effect of contamination from an M-type foreground star projected onto the galaxy, 
which could not be separated from the galaxy itself in the NIR images.\label{g1CCsed}}
\end{figure*}

\begin{figure*}
\centering
\hbox{
\includegraphics[width=0.5\linewidth,height=8cm]{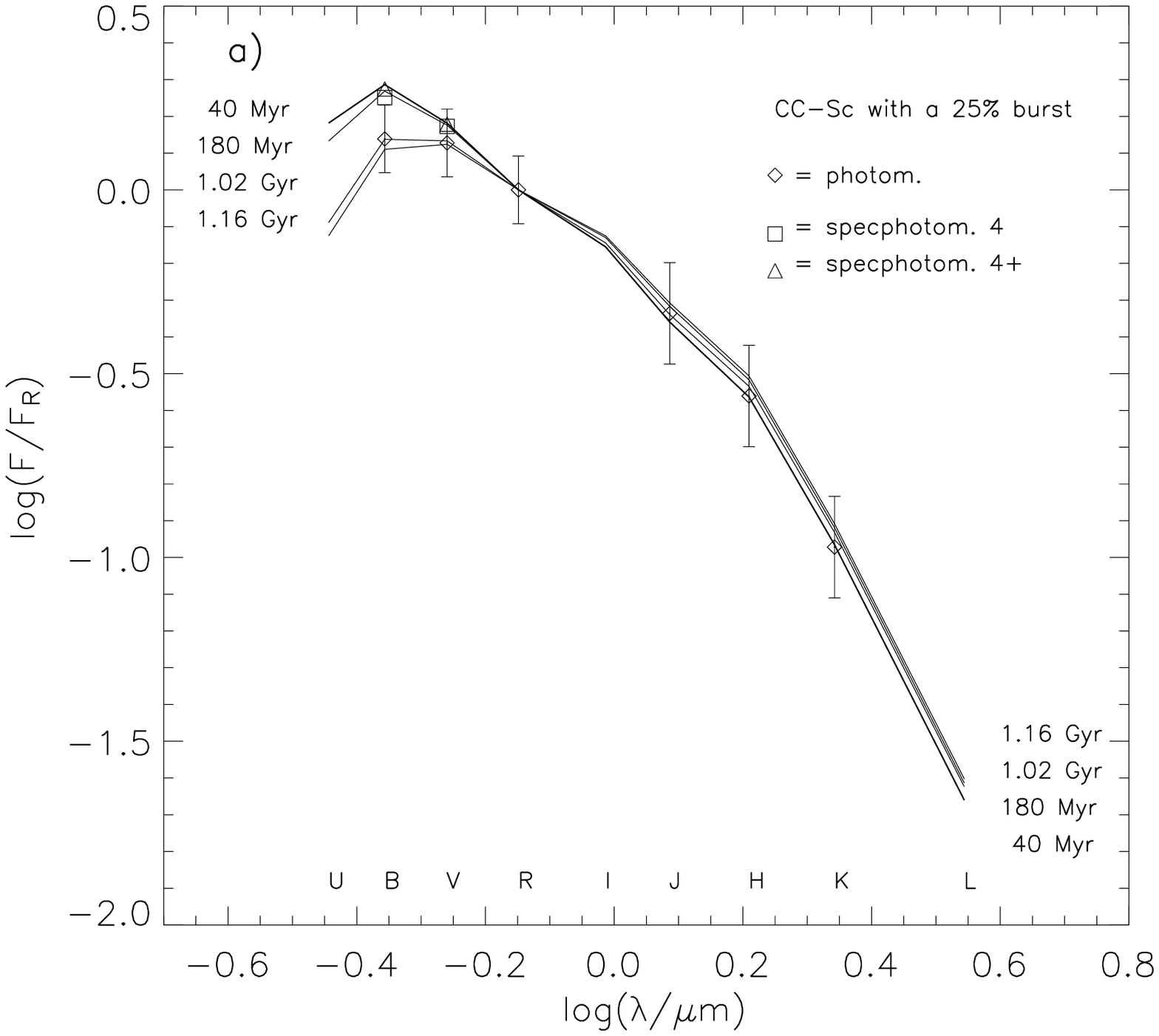}  
\includegraphics[width=0.5\linewidth,height=8cm]{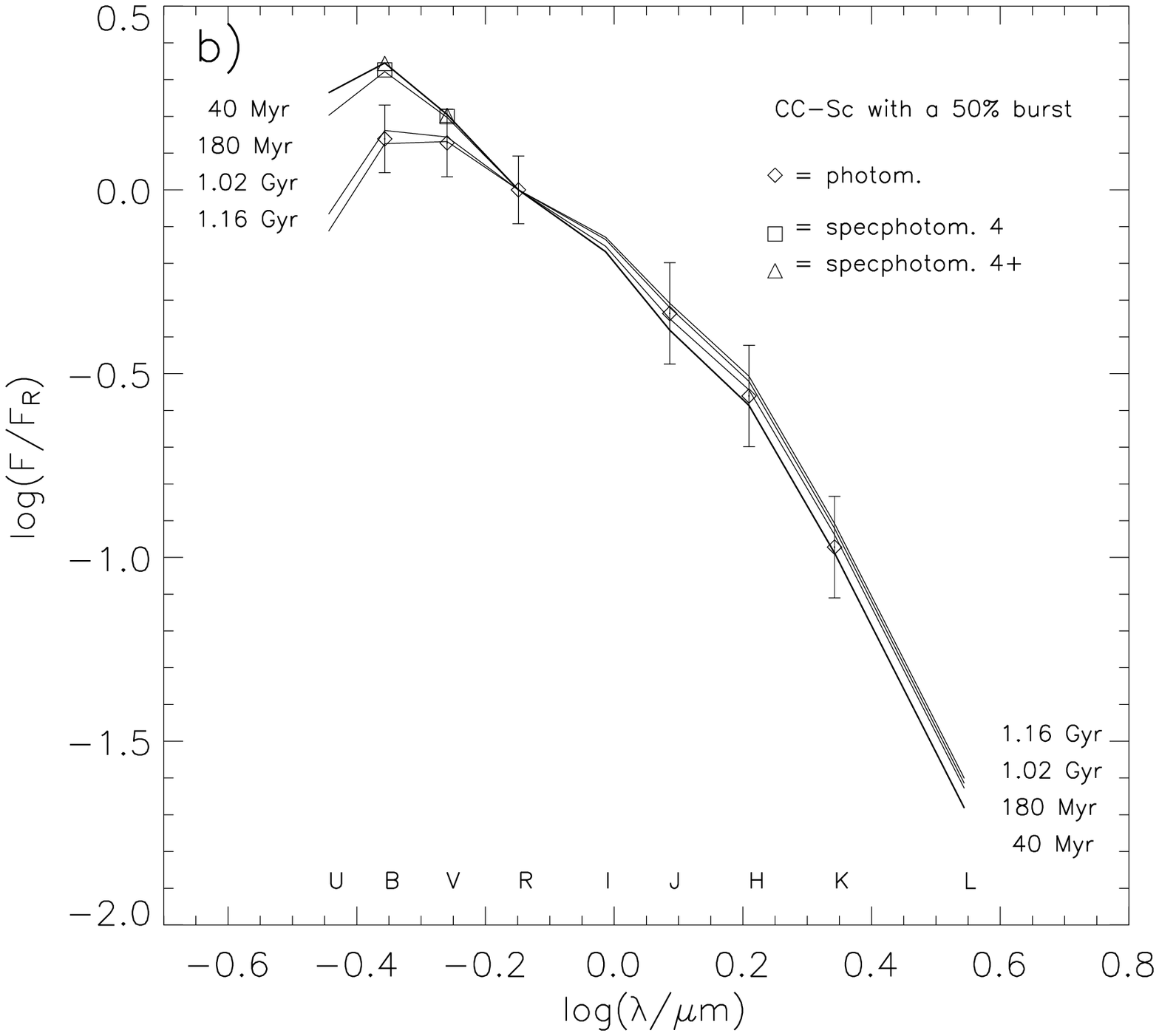}} 
\caption{SEDs for the CC-Sc models at differing ages after a weak [{\bf a)}] and a 
strong [{\bf b)}] burst. Photometric points of G4 are marked with diamonds, while
the spectrophotometric points derived from the dereddened spectra (according to
the best fit extinction values obtained from the spectral analysis) at 
P.A. 130$^{\circ}$ and 90$^{\circ}$ are indicated by triangles and squares, respectively. 
For models of both strengths, optical spectrophotometric points and NIR photometric points 
agree very well with a burst age in the range 40 to 180 Myr. The thick line indicates
the best fit age obtained from the spectra. \label{g4CCsed}} 
\end{figure*}

The optical-NIR spectral energy distribution (SED) served for a further comparison 
between models and observational data. 
SED diagrams \citep[see, e.g., ][]{krug95} were obtained from 
integrated broad-band fluxes convolved
with the Johnson filter response functions \citep{bess90} and normalized to the 
flux in the V (G1) or R (G4) band, as a function of the effective wavelength $\lambda_{\rm eff}$
of the filters. Flux values for the models, in units of erg cm$^{-2}$
s$^{-1}$ were obtained from the
absolute magnitudes calculated from the model spectra by a {\sc galev} sub-routine.
Flux values for the galaxies were obtained from the
measured magnitudes and, in the optical regime, also from their spectra. The available spectral 
ranges allowed us to derive $B$, $V$ fluxes for G1 and $B$, $V$, $R$ fluxes for G4. Photometric
zero points, as determined from the observed spectrophotometric standard stars, were included.
The diagrams
are shown in Figs.~\ref{g1CCsed}, \ref{g4CCsed} for the two CC-Sc models at several burst
ages. For G1, SEDs obtained from the strong burst model offer a better agreement with the 
data, while for G4 both burst strengths produce SEDs in good agreement with the
observations. Models
with a burst age in the range 40 to 180 Myr are favored, thus confirming the results from the 
above spectral analysis.

We note that for both galaxies the normalized spectrophotometric optical fluxes 
are slightly higher than the photometric values in the same bands ($B$ band for G1 and both 
$B$ and $V$ bands for G4) and are in very good agreement with the modeled SEDs (for strong burst).
We interpret this fact as a consequence of an insufficient correction of the observed magnitudes 
for internal reddening. 
While spectra have been corrected for internal reddening following the procedure described
above, only a standard inclination correction has been applied to the 
total magnitudes (Paper~II). 
The reason for this choice is that we are not able to determine with sufficient accuracy
the amount of internal extinction affecting the total magnitudes of the galaxies.
The dust distribution in G1 is inhomogeneous as shown by the differing color excesses measured
in spectra across distinct portions of the galaxy (E($B-V$)$_{\rm g}$ = 0.17, 0.35, 0.65 mag; 
see Paper II). A similar situation holds for G4 (see Paper IV), as already mentioned.
Hence, it is not clear which value of extinction correction is more suitable to be applied to
total magnitudes. By using as reference the average color excess measured from different spectra,
E($B-V$)$_{\rm g}$ = 0.5 mag for G1 and 0.6 mag for G4, we have shown in Fig.~\ref{colevolSc} how the
galaxies would be placed on a two-color diagram after correction for internal extinction,
under the assumption that the stellar continuum is affected by roughly half of the extinction 
affecting the emission lines \citep{cks94}. This is to be considered only an indication of the
effects of internal extinction on galaxy colors, thus we prefer not to use magnitudes and colors
corrected this way in the final comparison between observed and modeled quantities.
At least for G4 there is
observational indication (Paper~IV, Fig. 7) that an amount of extinction significantly higher 
than that assumed by the standard inclination correction is present all across the galaxy.
This is probably the case also for G1.
Since shorter wavelengths are more affected by extinction, the effects of an insufficient
correction are expected to be greater in $B$ than in $V$, as actually appears from
the comparison between photometric and spectrophotometric points in Fig.~\ref{g4CCsed}.

The normalized NIR observational fluxes of G1 appear higher than the modeled ones
(i.e. shifted toward older ages),
although still in agreement within the errors\footnote{The error bars in Fig.~\ref{g1CCsed}
represent only the error in the observed magnitudes (see Paper II for a discussion on 
photometric errors). Comparable errors have to be considered for the model SEDs. The errors
of modeled colors are estimated to be 0.05 mag in the optical and 0.1 mag in the NIR due to uncertainties in 
the isochrones and color calibrations.}.
This discrepancy in the NIR is most probably a consequence of contamination from the foreground star 
projected onto G1 (see Fig.~\ref{RGB}), which displays a typical M-type spectrum and could not 
be separated from the galaxy itself when doing NIR photometry.

Observational and modeled colors and metallicities
are shown in Table~\ref{colorsg1g4} for CC-Sc models with different burst strengths
and ages. 
In Table~\ref{lumg1g4} absolute magnitudes of G1 and G4 (distance modulus ${\rm m - M =}$ 36.276)
are compared with model ones, calculated for burst strengths $b$ = 0.5 and $b$ = 0.25.
Model luminosities, masses, and SFRs have been rescaled taking as reference the
R-band luminosity of G1 and the K-band luminosity of G4.
Once again we stress that a standard inclination correction has been applied to observed
colors and magnitudes to account for internal extinction, but this correction was probably
insufficient as suggested by the Balmer decrement measured in the individual spectra.
In fact, the discrepancies between photometric measurements in the $B$ and $V$ bands 
and the model SEDs visible in Figs.~\ref{g1CCsed}, \ref{g4CCsed} are found also in the 
direct comparison between observed and modeled colors and luminosities.
Taking into account this fact, the comparison between observed and model luminosities
indicates that G1 is roughly consistent with a total mass (star + gas) in the
range 4 - 6$\times$10$^{9}$ M$_{\odot}$ (using as reference the 
strong burst models)
and G4 has luminosities in good agreement with 
$\sim$ 2 - 3$\times$10$^{10}$ M$_{\odot}$ models 40 to 180 Myr after a moderate to strong burst. 
The modeled stellar mass-to-light ratios in the K band are 0.62 (0.65) and 
0.46 (0.55) M$_{\odot}$ L$^{-1}_{\odot}$
for a 40 and a 180 Myr old strong (moderate) burst, respectively.
The metallicities obtained with chemically consistent models are comparable to (or slightly higher 
than) those determined from the observed emission-line ratios.
SFRs obtained from models with burst age 40 Myr
(taking into account the assumed FVM = 0.5, see Table~\ref{lumg1g4}) are in reasonable 
agreement with measurements for G1. In the case of G4, models with SFRs in better agreement
with the observations are those with burst age 180 Myr, while models with younger age give
too high a SFR. As already mentioned in Sect.~2.1 and in Paper IV, the internal extinction map of G4 
strongly suggests that the SFR derived from the H$\alpha$ emission-line is 
a lower limit to the actual SFR and continuum radio observations suggest that the actual 
SFR of G4 is $\ga$ 6 M$_{\odot}$ yr$^{-1}$, i.e. close to the SFR for the strong burst model
with age 180 Myr.

In the strong burst model, the mass percentage of stars produced since the onset of the burst is
$\approx$ 9\% and 24\% after 40 and 180 Myr, respectively, and they contribute 44\% and 63\% of the
visual light. 


\begin{figure}[ht]
\centering
\includegraphics[width=\linewidth]{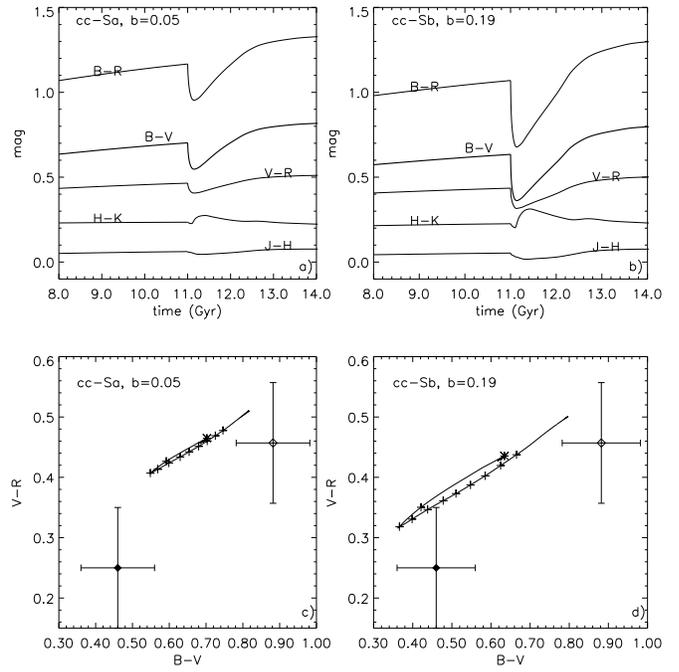}  
\caption{{\bf a)} Evolution of color indices for a cc-Sa model with a 5\% burst 
switched on after 11 Gyr of undisturbed evolution. The $B-V$ index is indicated with a thick 
line to avoid confusion with other indices. For the same reason, an offset of $-$0.5
has been applied to the $J-H$ index. {\bf b)} Same as {\bf a)}, for cc-Sb model 
with a 19\% burst.
{\bf c)} Color-color diagram showing the color indices evolution during the burst phase for
a cc-Sa model with a 5\% burst. The asterisk marks the onset of the burst at a galaxy age of
11 Gyr. Plus symbols are plotted for burst ages from 40 Myr to 1.28 Gyr at intervals of 140 Myr.
The empty diamond indicates the observed colors of G2, after a standard
inclination correction. The full diamond indicates the observed colors after correction for
internal extinction using E($B-V$)$_{\rm g}$ = 0.52 mag, as derived from the Balmer decrement.
{\bf d)} Same as {\bf c)} for a cc-Sb model with a 19\% burst. \label{colevolSaSb}}
\end{figure}

\section{Early-type group member: G2}

Both photometric and spectroscopic results (Paper~II) have shown that 
G2 is a strongly reddened early-type galaxy with evidence of present-day 
star formation activity at its center.
The Balmer decrement in its spectrum implies an internal extinction A$_{V}$ = 1.6 mag.
Although such a high extinction value is probably mostly due to dust associated with 
the central star formation, it is possible that the inclination correction we applied to
the total magnitudes (A$_{V}$ = 0.8, see Paper~II) was insufficient to account for the amount of
extinction affecting this galaxy, as in the case of the two spiral galaxies discussed above.
Nevertheless, we preferred to maintain the standard inclination correction concerning
total magnitudes because it is unclear weather the extinction value derived from the
long-slit spectrum can be extended to the whole galaxy. Additionally, due to the
low signal-to-noise ratio of the H$\beta$ emission line, the uncertainty in the derived extinction
in quite high ($\pm$ 1.2 mag).

The strong Balmer absorption lines (in which the emission lines are embedded)
and the CaII absorption index suggest that the spectrum is dominated by a population 
of A-type stars. The emission line ratios are consistent with a subsolar
metallicity of the gaseous component (Z $\la$ 0.5 Z$_{\odot}$).

In Paper IV (Sect. 7.1) we have proposed two possible explanations for the observed properties of G2:
either this galaxy is the outcome of a merger that induced a (now fading) burst of star
formation and produced part of the faint tails and plumes visible in the group, or it 
is an early-type galaxy which started a small episode of star formation as a consequence
of tidal interaction with the companion spiral galaxies. In fact, several numerical
simulations have shown that interactions and merger events may trigger gas inflows and
subsequent central star formation \citep[e.g. ][]{no88,bh91,mbr93,mh94,bh96}.
Here we use evolutionary synthesis models to investigate both possibilities.
To test the first possibility, we simulate a burst of star formation induced by the
merger of two early-type spiral galaxies, i.e. two Sa or Sb galaxies (at the moment
{\sc galev} does not support the simulation of mixed-morphology mergers). We investigate the
second possibility using the model of an elliptical (E) galaxy on which we turn on a burst
of star formation at late ages.

\subsection{Early type spiral-spiral merger}

\begin{figure*}
\centering
\hbox{
\includegraphics[width=0.5\linewidth]{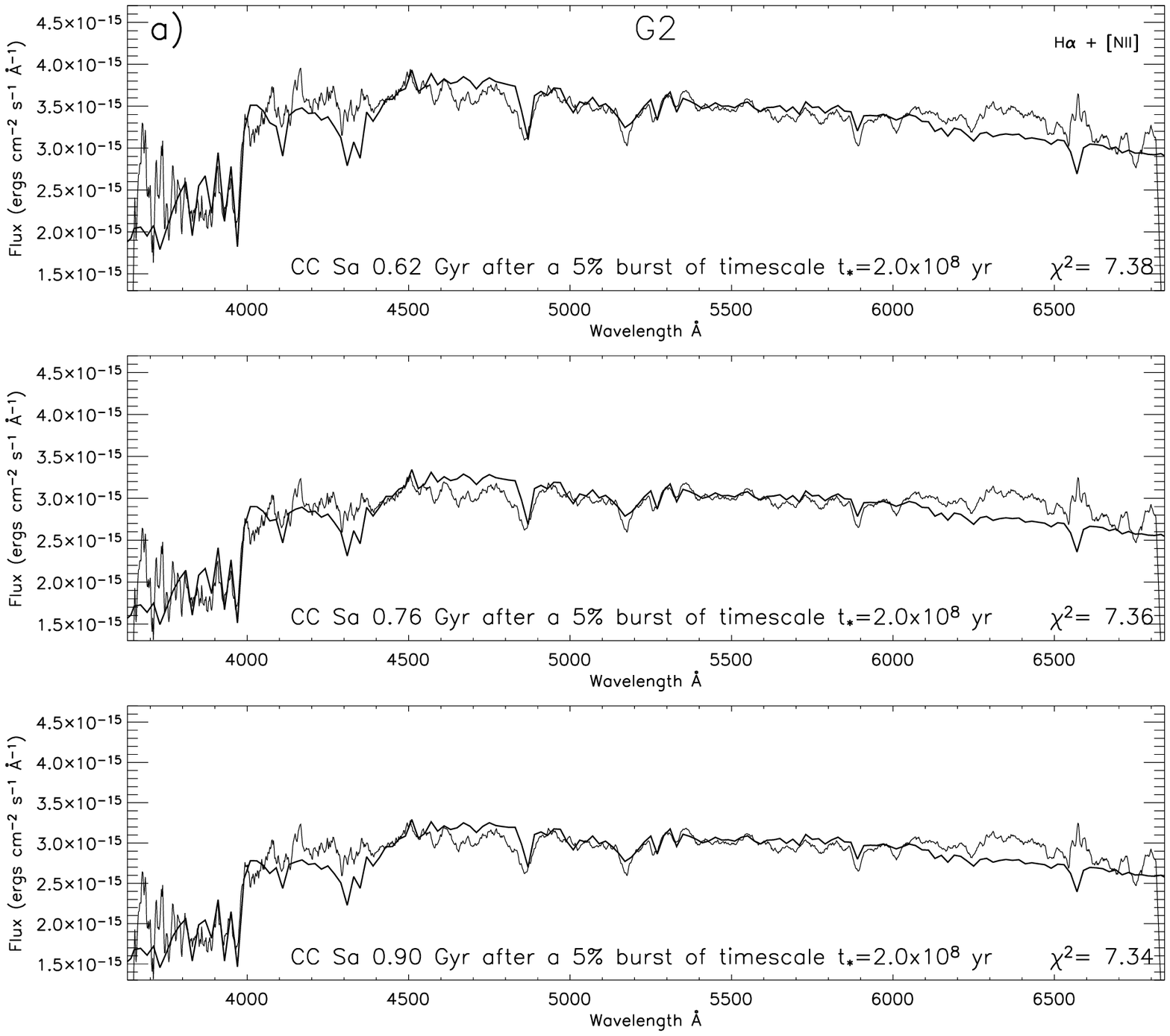}  
\includegraphics[width=0.5\linewidth]{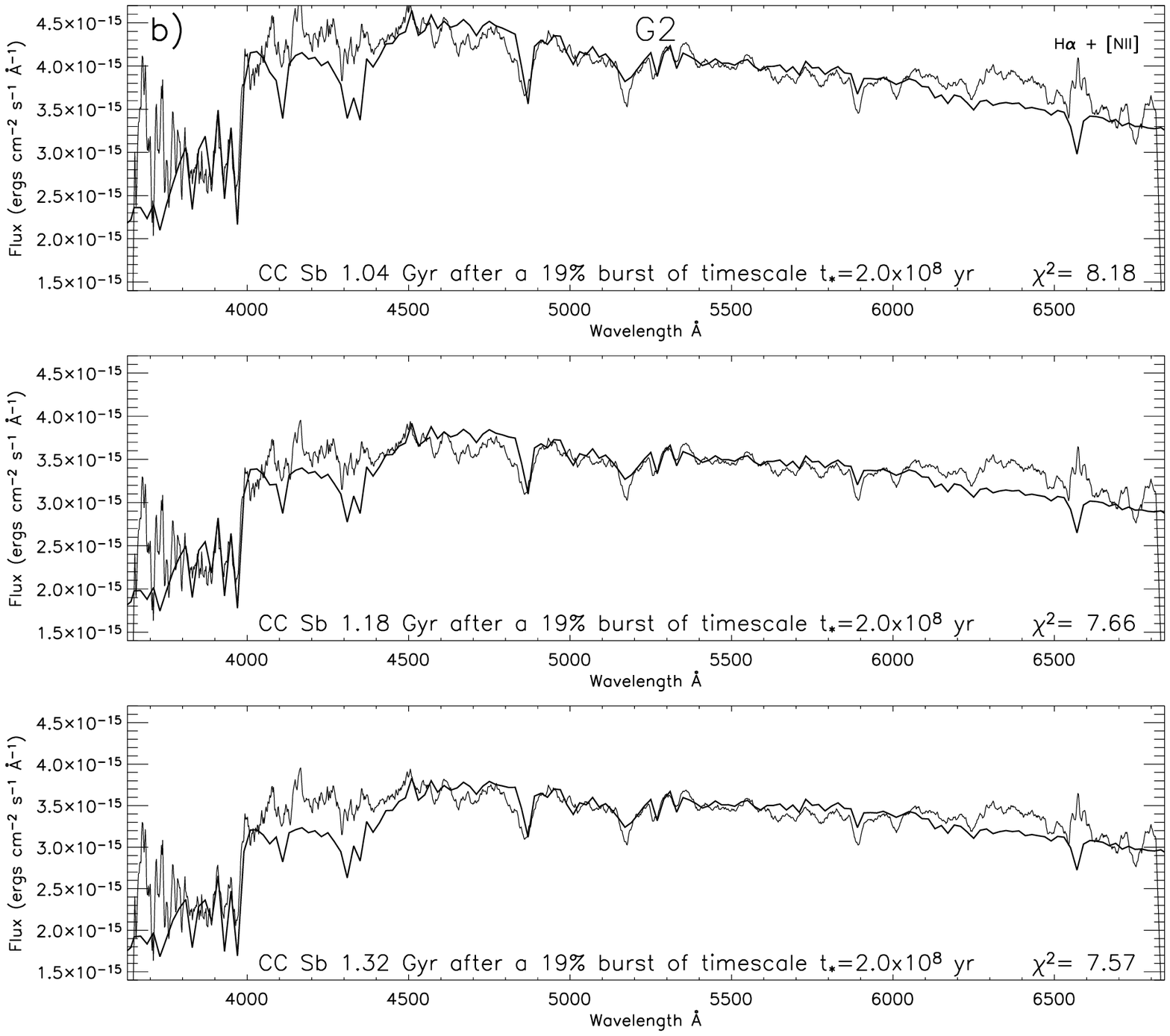}} 
\caption{{\bf a)} CC-Sa model spectra (thick line), conveniently scaled,
overplotted to the dereddened and
smoothed observed spectrum of G2 at three different ages after
the onset of a 5\% burst. The minimum reduced $\chi^2$ is obtained for 
E($B-V$) = 0.36 mag and burst age 900 Myr, however the other two models
shown give nearly as good a result with E($B-V$) = 0.41 and 0.36 mag at
ages 620 and 760 Myr, respectively. The fit worsens sensibly at younger/older
ages than displayed.
{\bf b)} Same as {\bf a)} for a CC-Sb model at three different ages after
the onset of a 19\% burst. The best fit occurs for E($B-V$) = 0.41 mag and 
burst ages between 1.2 and 1.3 Gyr.} 
\label{SaCCspec}
\end{figure*}

We compared the observational data with two models in which the 
progenitors are two Sa or Sb galaxies. 
A burst of star formation with exponential decay time-scale t$_{\ast}$ 
= 2.0$\times$10$^8$ yr was turned on when the galaxies were 11 Gyr old\footnote{Turning 
on the burst in a later epoch (e.g. 12 Gyr) would not change
significantly the results.}.
The color evolution for these two models and an example of color loop in a color-color
diagram during the burst phase are shown in Fig.~\ref{colevolSaSb}.
The position of G2 in the color-color diagrams is shown after a standard
inclination correction (empty diamond) and as it would appear after a correction 
for internal extinction using the color excess 
derived from the Balmer decrement, E$(B-V)_{\rm l}$ = 0.52 mag (but see the above
remarks concerning the application of the extinction correction to total magnitudes).
Photometric data appear consistent with an already 
aging burst (age $>$ 0.5 Gyr), as do the A-star spectral features.

In the Sa model only a weak burst, with maximum strength $\sim$ 5\%, 
can be simulated due to the limited amount of gas remaining in 
the galaxy after 11 Gyr of undisturbed evolution. According to the CaII
absorption index, the model spectra $<$ 0.5 Gyr after the burst
show too young a stellar population with respect to the observations.
Unlike the case of the two spiral galaxies, we did not take 0.5$\times$E($B-V$)$_{\rm g}$ 
as initial guess for the internal extinction in the continuum, E($B-V$),
because this relation was derived by \citet{cks94} for starburst galaxies, which
G2 is not (anymore). However, since not only the observed spectrum but also relative photometry
suggested a strong reddening, we used the value determined from
the Balmer decrement as an upper limit to E($B-V$) and we applied an analogous fitting 
procedure to that described in Sect. 3 to determine the optimal extinction value 
together with the best fit burst age. The value of E($B-V$) derived this way is 
$\approx$ 0.4 mag (specifically in the range 0.36 - 0.41 mag), which is
somewhat lower than the measured E($B-V$)$_{\rm g}$, although 
well within the error of the measurement.
The best agreement with the extinction corrected, smoothed spectrum
is found 900 Myr after the onset of the burst (minimum reduced-$\chi^2$ = 7.34),
however also model spectra at burst ages 620 and 760 Myr give a nearly as good
representation of the observed spectrum (reduced-$\chi^2$ = 7.38 and 7.36, 
respectively; Fig.~\ref{SaCCspec}).  We note that the H$\beta$ absorption line
becomes weaker at ages $\ga$ 760 Myr, while it better matches the observed feature
at a younger burst age.

In the Sb model, at the same galaxy age, the remaining gas content is higher and
a stronger burst can be turned on. We considered a 19\% 
burst turned on at a galaxy age of 11 Gyr. In this case the signatures of A-type
stellar populations
remain visible for a longer time after the start of the burst and begin fading
only after $\ga$ 1.3 Gyr, when the Balmer absorption lines weaken and the CaII 
absorption index increases toward unity. The best agreement with the observed spectrum
is found $\sim$ 1.2 to 1.3 Gyr after the onset of the burst (Fig.~\ref{SaCCspec}).

\begin{figure*}
\centering
\hbox{
\includegraphics[width=0.5\linewidth]{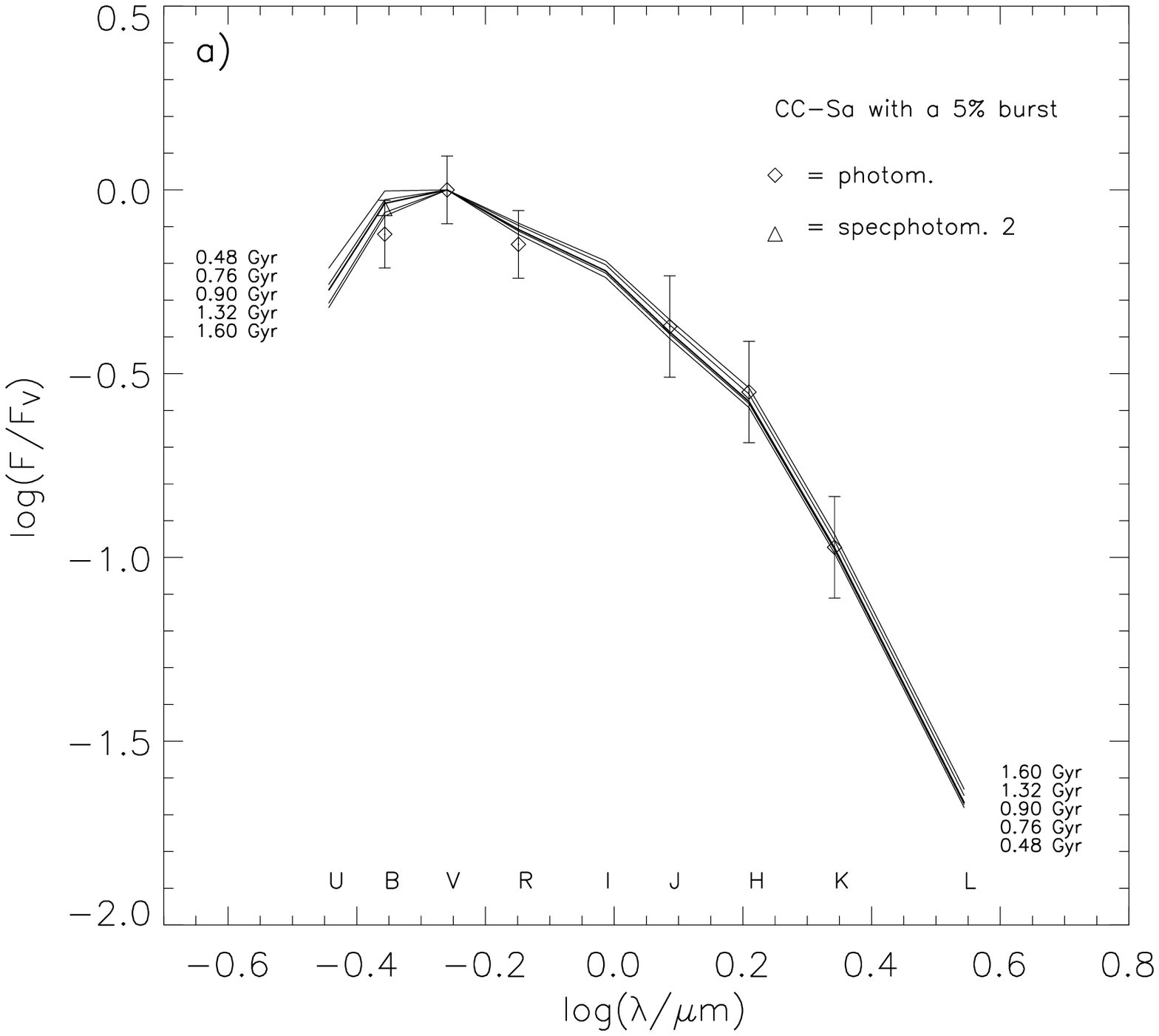} 
\includegraphics[width=0.5\linewidth]{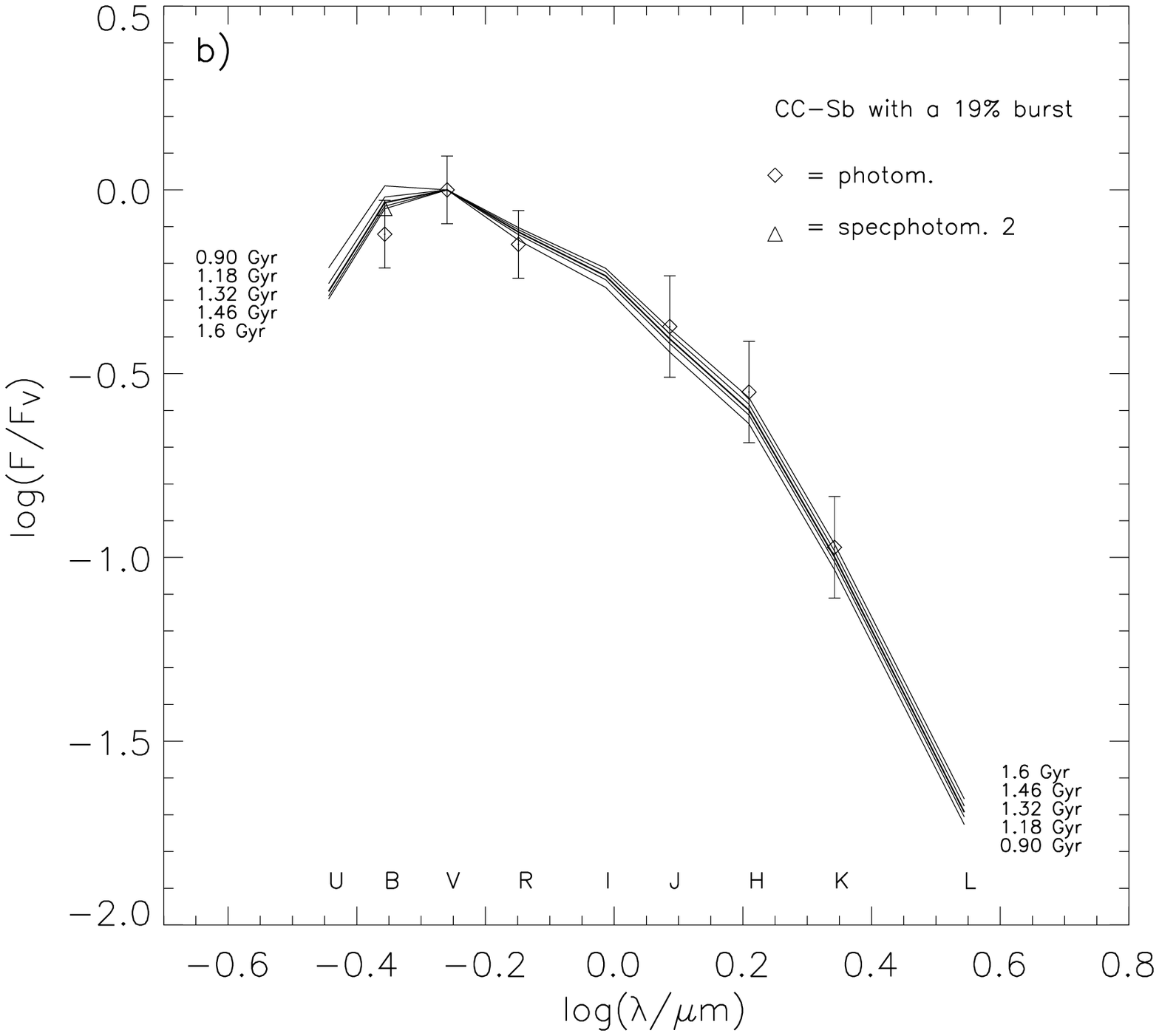}} 
\caption{SEDs calculated for a wide range of burst ages for the CC-Sa [{\bf a)}] and Sb
[{\bf b)}] models along with dereddened photometric and spectrophotometric 
data points for G2. The thick lines indicate the best-fit models 
according to the spectral analysis.} 
\label{g2sed}
\end{figure*}

\begin{table*}
\caption[]{G2: Observed$^{\mathrm{a}}$ and Modeled Colors}
\label{compg2}
$$
\begin{tabular}{llllllllll}
\hline
\hline
\noalign{\smallskip}
Object/Model & $B-V$ & $V-R$ & $V-K$ & $B-R$ & $B-H$ & $J-H$ & $H-K$ & Z/Z$_{\odot}$ \\
\noalign{\smallskip}
\hline
\noalign{\smallskip}
G2 & 0.88 & 0.46 & 2.52 & 1.34 & 3.24 & 0.59 & 0.16 & $\sim$0.5\\
 & & & & & & & & \\
CC-Sa, M = 2.5$\times$10$^{10}$ M$_{\odot}$, b=0.05 & & & & & & &  & \\
 480 Myr &0.59  &0.52  & 2.48 &1.11  & 2.85 & 0.57 & 0.22 & 0.68 \\ 
 620 Myr &0.62  &0.54  & 2.49 &1.15  & 2.89 & 0.57 & 0.22 & 0.67 \\
 760 Myr &0.65  &0.55  & 2.50 &1.19  & 2.93 & 0.57 & 0.22 & 0.64 \\
 900 Myr &0.67  &0.56  & 2.51 &1.23  & 2.97 & 0.56 & 0.22 & 0.60 \\
 & & & & & & & & \\
CC-Sb, M = 2.5$\times$10$^{10}$ M$_{\odot}$, b=0.19 & & & & & & & &  \\  
 900 Myr & 0.55 & 0.49 & 2.36 & 1.04 & 2.70 & 0.55 & 0.22 & 0.64\\
1.04 Gyr & 0.59 & 0.50 & 2.39 & 1.09 & 2.76 & 0.55 & 0.22 & 0.61\\ 
1.18 Gyr & 0.63 & 0.52 & 2.42 & 1.15 & 2.83 & 0.55 & 0.21 & 0.59\\
1.32 Gyr & 0.67 & 0.54 & 2.45 & 1.21 & 2.91 & 0.55 & 0.21 & 0.57\\
\noalign{\smallskip}
\hline
\end{tabular}
$$
\begin{list}{}{}
\item[$^{\mathrm{a}}$] Observed colors after the application of a standard inclination correction.
\end{list}
\end{table*}

Both models, obtained considering a chemically
consistent evolution, provide a satisfactory match to the
metallicity-sensitive iron lines like Fe $\lambda$5015, CaFe $\lambda$5269, Fe
$\lambda$5406, Fe $\lambda$5709, and Fe $\lambda$5782 \citep{wor94}.
The gas metallicity reached by the Sb model 
at the best-fit age is Z = 0.0106 $\sim$ 0.59 Z$_{\odot}$ 
in good agreement with the metallicity derived in Paper~II from the emission-line
ratios, whereas it is slightly higher for the Sa model (Table~\ref{compg2}). 

In Table~\ref{compg2} the observed, face-on colors are compared with colors
from the two models at four different burst ages each. 
The color comparison favors the older burst models, in particular the Sa model
at burst ages of 760 - 900 Myr. Significant discrepancies are found in the
colors involving the $B$ band. In fact in all models these are bluer than
observed. In connection with this fact, we note that on the SED diagrams in
Fig.~\ref{g2sed} the spectrophotometric flux in the $B$ band is higher than
the photometric one, like in the case of G1 and G4. The normalized $B$ band 
spectrophotometric flux and the photometric fluxes in the other bands appear
largely in agreement with the model SEDs. Once again, the Sa model is
the one better approaching the observations. The discrepancy in the $B$ band
would confirm an insufficient extinction correction of the observed magnitude,
particularly visible in $B$, where extinction effects are stronger.

\begin{table*}
\caption[]{G2: Observed$^{\mathrm{a}}$ and Modeled Luminosities}
\label{lumg2}
$$
\begin{tabular}{llllllll}
\hline
\hline
\noalign{\smallskip}
Object/Model & M$_{B}$ &  M$_{V}$ & M$_{R}$ & M$_{J}$ & M$_{H}$ & M$_{K}$ & SFR (M$_{\sun}$ yr$^{-1}$)\\
\noalign{\smallskip}
\hline
\noalign{\smallskip}
G2 & $-$19.82 & $-$20.70 & $-$21.16 & $-$22.47 & $-$23.06 & $-$23.22 & 0.4 - 0.8\\
 & & & & & & & \\
CC-Sa, b=0.05& & & & & & & \\
 480 Myr, M = 3.19$\times$10$^{10}$ M$_{\odot}$ & $-$20.15 & $-$20.74 & $-$21.27 & $-$22.43 & $-$23.00 &$-$23.22 & 0.44\\ 
 620 Myr, M = 3.28$\times$10$^{10}$ M$_{\odot}$ & $-$20.11 & $-$20.73 & $-$21.26 & $-$22.43 & $-$23.00 &$-$23.22 & 0.22 \\
 760 Myr, M = 3.37$\times$10$^{10}$ M$_{\odot}$ & $-$20.07 & $-$20.72 & $-$21.26 & $-$22.43 & $-$23.00 &$-$23.22 & 0.12\\
 900 Myr, M = 3.45$\times$10$^{10}$ M$_{\odot}$ & $-$20.03 & $-$20.71 & $-$21.26 & $-$22.44 & $-$23.00 & $-$23.22 &0.06 \\
 & & & & & & & \\
CC-Sb, b=0.19 & & & & & & & \\
 900 Myr, M = 2.72$\times$10$^{10}$ M$_{\odot}$ & $-$20.30 & $-$20.86 & $-$21.34 & $-$22.45 & $-$23.00 &$-$23.22 & 0.14\\
1.04 Gyr, M = 2.77$\times$10$^{10}$ M$_{\odot}$ & $-$20.24 & $-$20.83 & $-$21.34 & $-$22.46 & $-$23.00 &$-$23.22 & 0.07\\ 
1.18 Gyr, M = 2.75$\times$10$^{10}$ M$_{\odot}$ & $-$20.17 & $-$20.80 & $-$21.32 & $-$22.45 & $-$23.00 &$-$23.22 & 0.04\\
1.32 Gyr, M = 2.69$\times$10$^{10}$ M$_{\odot}$ & $-$20.10 & $-$20.77 & $-$21.31 & $-$22.45 & $-$23.00 &$-$23.22 & 0.02\\
\noalign{\smallskip}
\hline
\end{tabular}
$$
\begin{list}{}{}
\item[$^{\mathrm{a}}$] Observed luminosities after the application of a standard inclination correction.
\end{list}
\end{table*}

Model luminosities and total (star + gas) masses, rescaled to the K-band observed
luminosity, are listed in Table~\ref{lumg2} together with observed luminosities. 
Except the $B$-band, where the measured luminosity is
probably underestimated, the older burst ages offer a reasonable agreement with 
the observations, especially for the Sa model. 
The total mass of G2 implied by the Sa and Sb models would be $\sim$ 3.4$\times$10$^{10}$ and
2.7$\times$10$^{10}$ M$_{\sun}$, respectively. This is equivalent to assume an
equal mass merger of two Sa (Sb) galaxies of mass 1.7$\times$10$^{10}$ (1.4$\times$10$^{10}$)
M$_{\sun}$ each, which led to the observed S0 morphology.
Predicted SFRs rescaled to the above total masses are
given in Table~\ref{lumg2}.
The agreement with observations -- taking into 
account the errors in the observations and in the models -- is reasonable for the Sa model,
while SFRs predicted by the Sb model are too low compared to the measured values.

According to the Sa (Sb) model, the stellar mass of G2 is  3.4$\times$10$^{10}$ 
(2.5$\times$10$^{10}$) M$_{\sun}$,
5\% (11\%) of which has been produced in a $\sim$ 0.7 - 0.9 (1.2 - 1.3) Gyr 
old burst. Stars produced during the burst contribute $\sim$ 6 - 3\% ( 10 - 8\%) 
of the visual light.

\subsection{Burst in an elliptical galaxy}


\begin{figure}[ht]
\centering
\includegraphics[width=\linewidth]{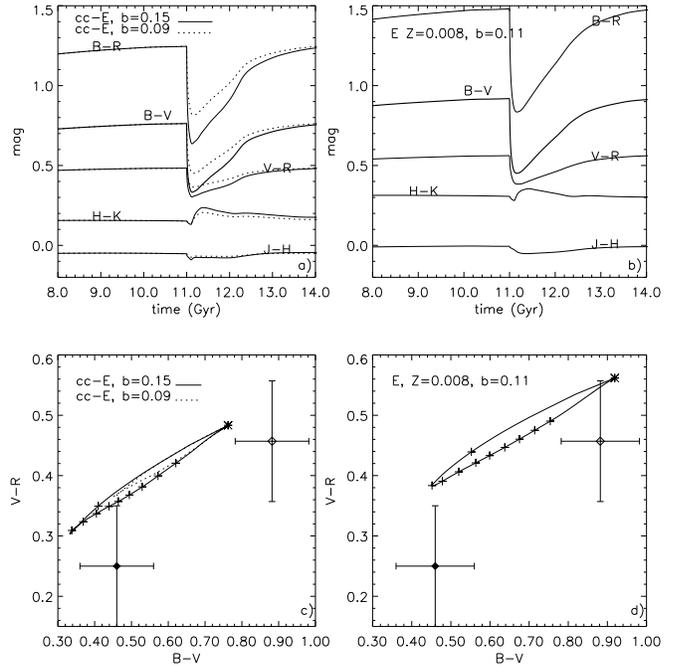}  
\caption{{\bf a)} Evolution of color indices for a CC-E model with a 15\% (solid line) and a
9\% (dotted line) burst switched on after 11 Gyr of undisturbed evolution. 
The $B-V$ index is indicated with a thick line to avoid
confusion with other indices. For the same reason an offset of $-$0.6 has been applied to
the $J-H$ index. {\bf b)} Same as {\bf a)}, for a non-CC E model 
with metallicity Z = 0.008 and a 11\% burst.
{\bf c)} Color-color diagram showing the color indices evolution during the burst phase for
the CC-E model. Line styles are as in {\bf a)}. Symbols as in Fig.~\ref{colevolSaSb}.
{\bf d)} Same as {\bf c)} for the non-CC E model. \label{colevolE}}
\end{figure}

\begin{figure*}
\centering
\hbox{
\includegraphics[width=0.5\linewidth]{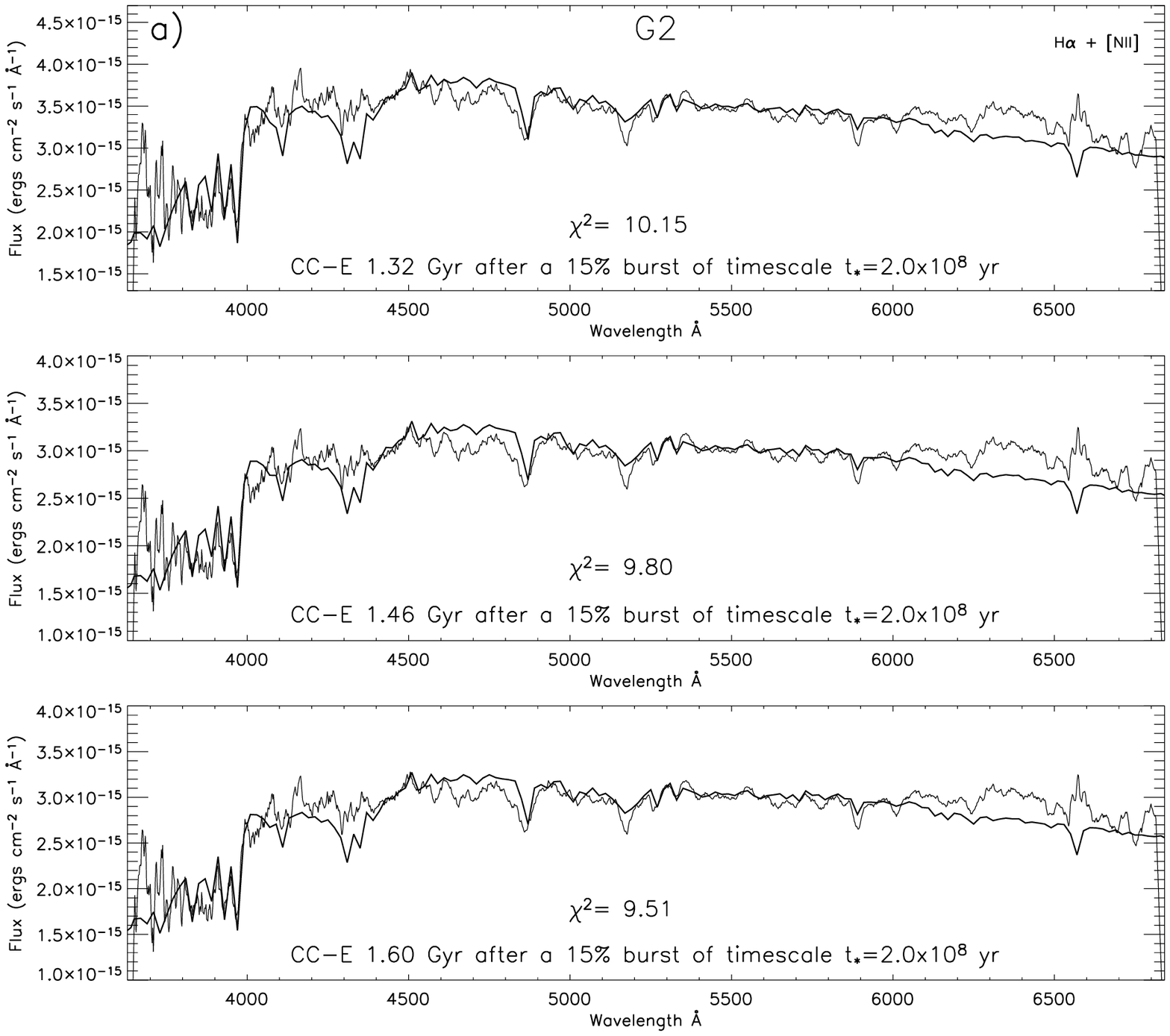}  
\includegraphics[width=0.5\linewidth]{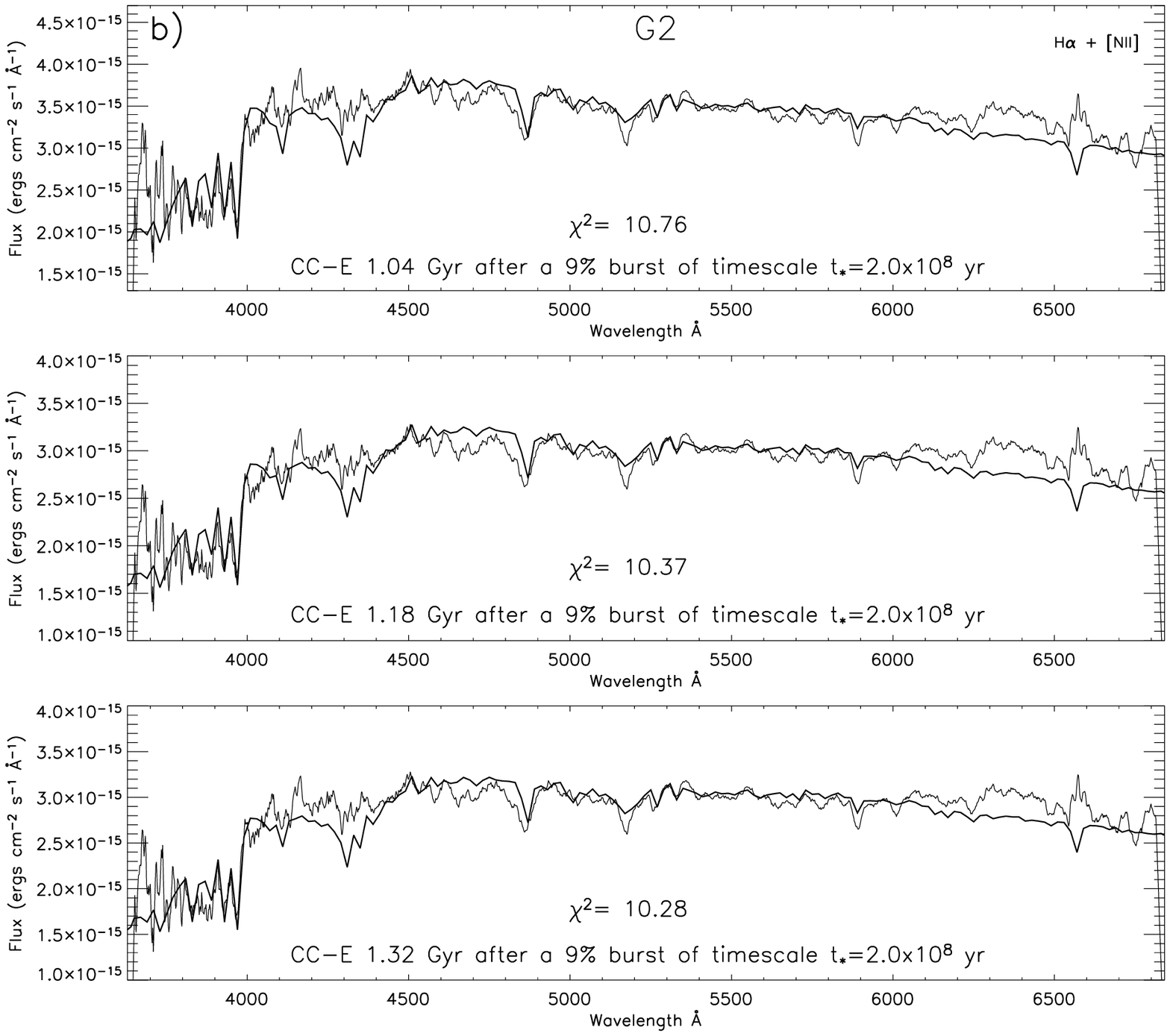}}  
\caption{{\bf a)} CC-E model spectra (thick line), conveniently scaled,
overplotted to the dereddened and
smoothed observed spectrum of G2 at three different ages after
the onset of a 15\% burst with a timescale of 2$\times$10$^8$ yr. 
The best fit occurs 1.5 to 1.6 Gyr after the burst, for an
internal extinction E($B-V$) = 0.36 mag.
{\bf b)} Same as {\bf a)} for a 9\% burst. The best fit occurs 1.3 
to 1.5 Gyr after the burst, with internal extinction E($B-V$) = 0.36 mag.} 
\label{g2Eccspec}
\end{figure*}

\begin{figure*}
\centering
\hbox{
\includegraphics[width=0.5\linewidth]{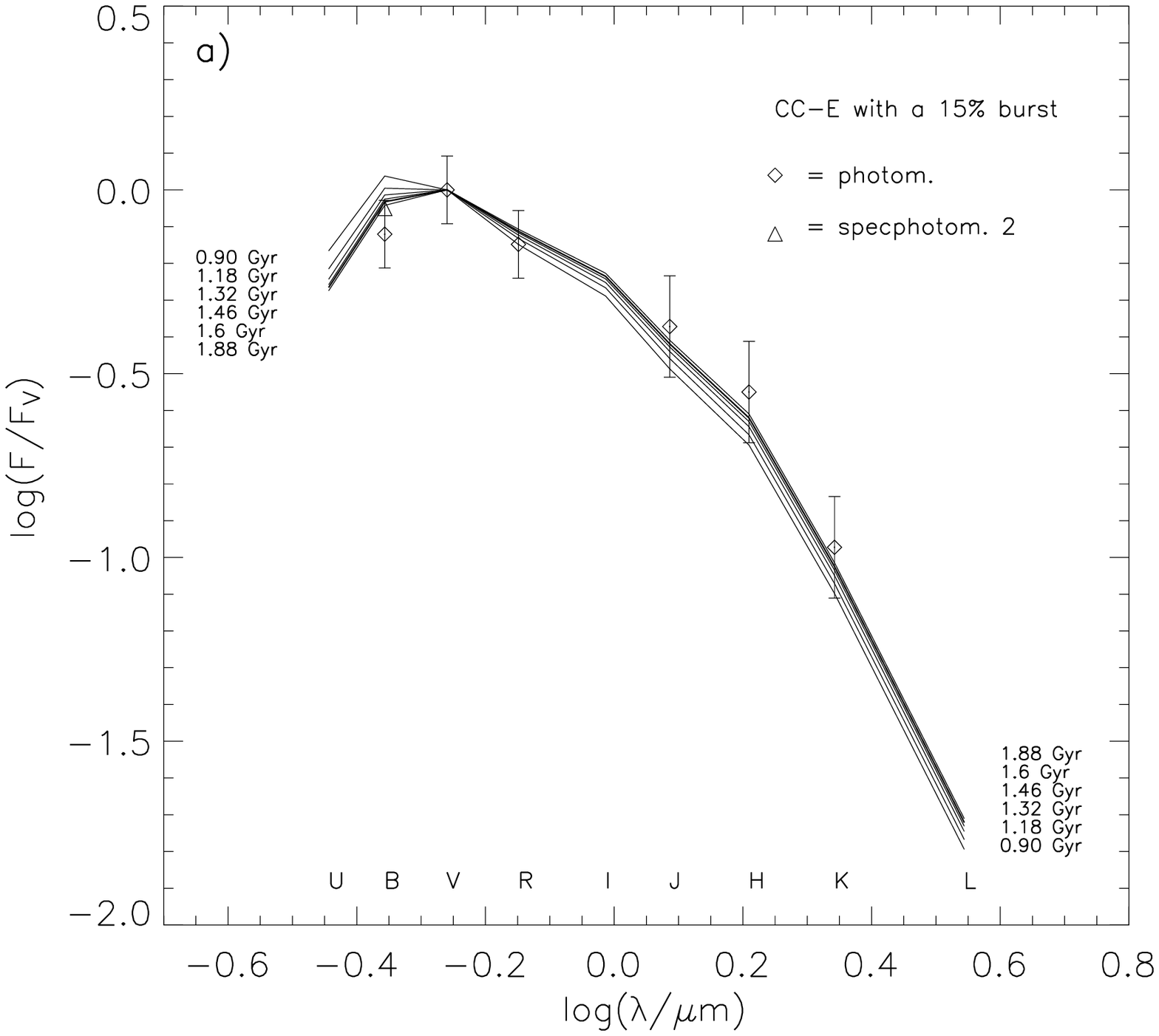} 
\includegraphics[width=0.5\linewidth]{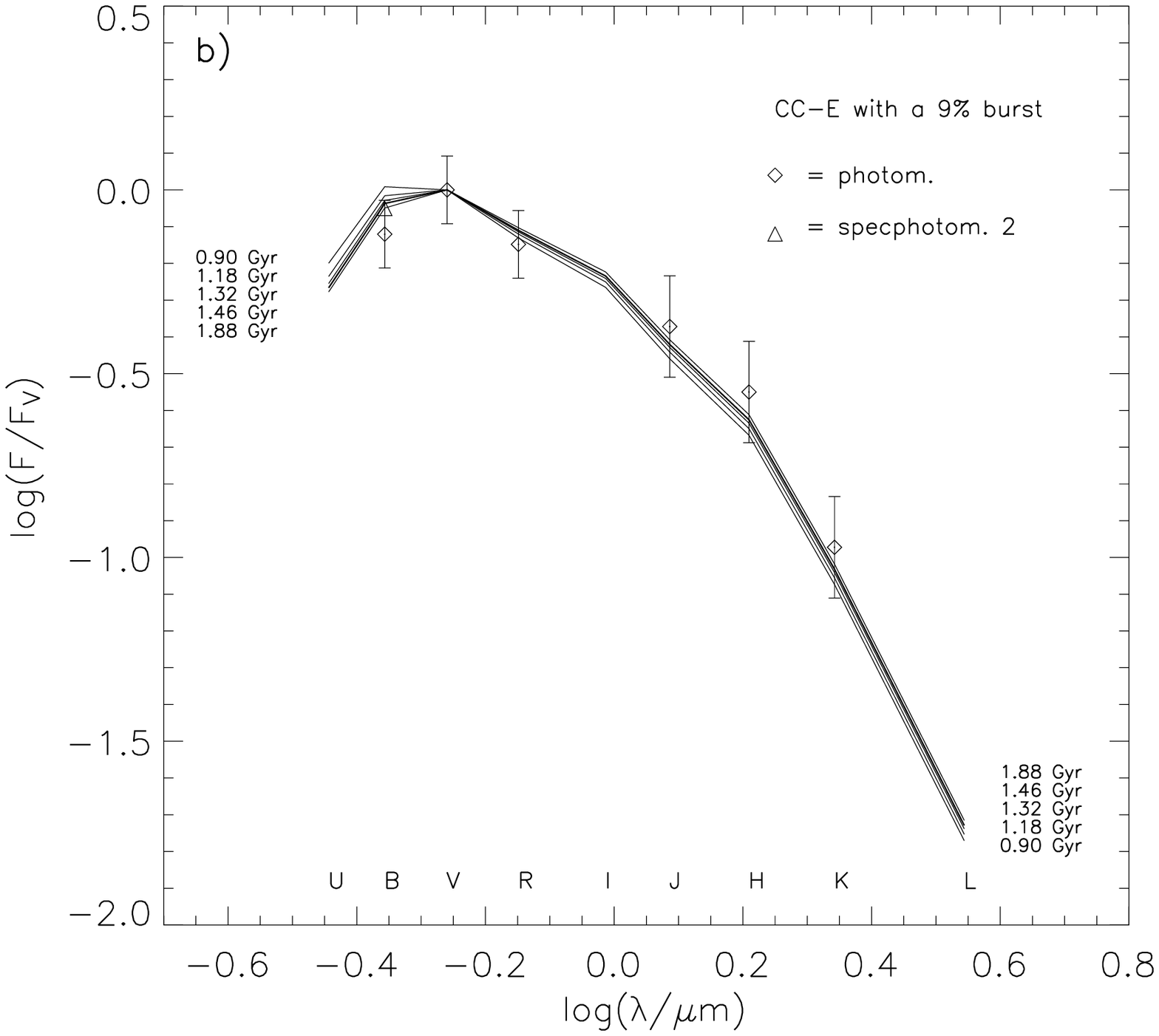}} 
\caption{SEDs for the CC-E models with burst strength 15\% (\emph{left}) and 9\%
(\emph{right}) along with dereddened photometric (diamonds) and spectrophotometric (triangle)
data points for G2. Best fit ages according to the spectral analysis are indicated with
a thick line. Discrepancies are visible in the NIR.} 
\label{g2Eccsed}
\end{figure*}

\begin{figure*}
\centering
\hbox{
\includegraphics[width=0.55\linewidth,height=6cm]{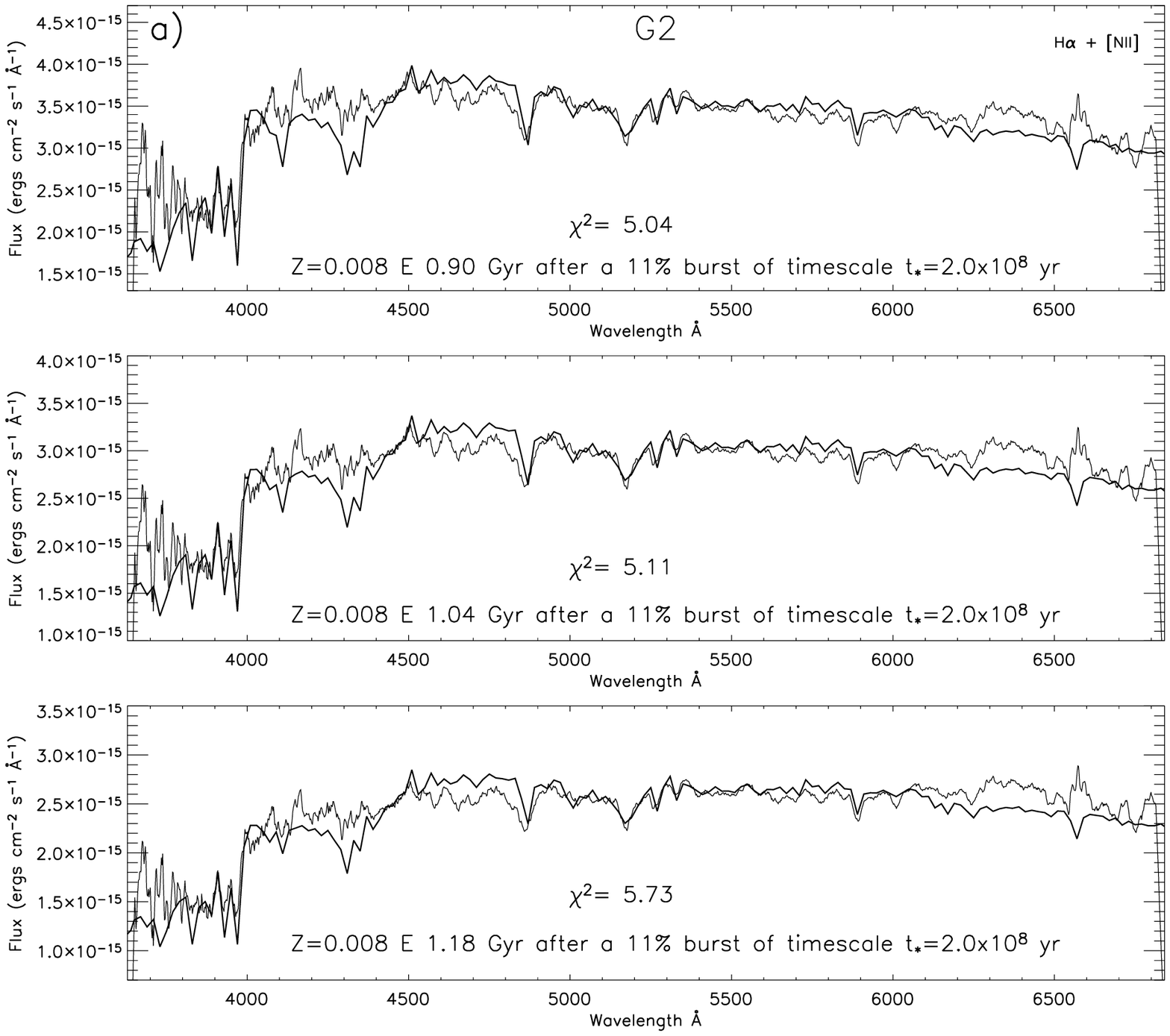} 
\includegraphics[width=0.45\linewidth,height=6cm]{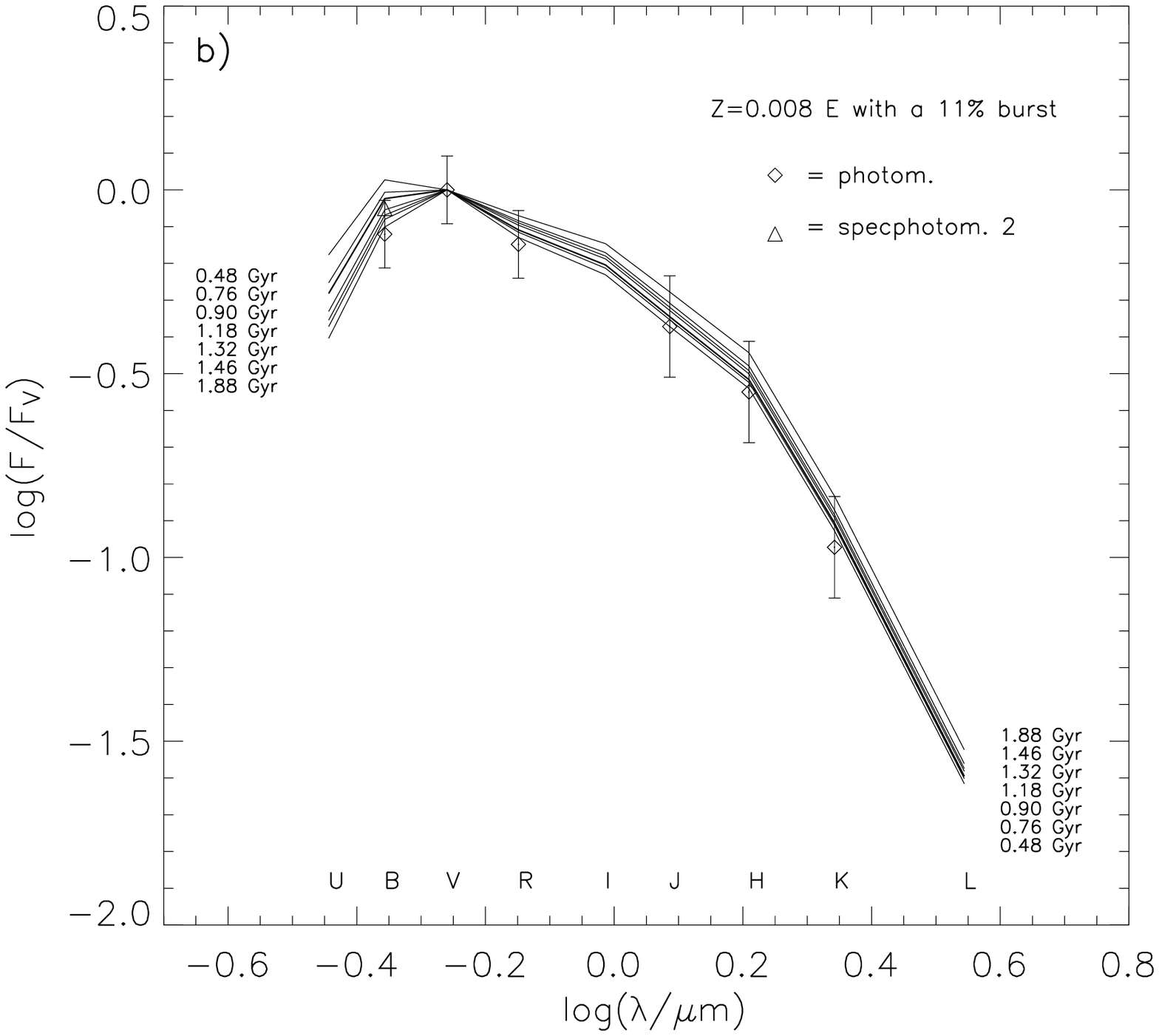}} 
\caption{{\bf a)} Smoothed spectrum of G2 corrected for internal reddening and overlapped with 
conveniently rescaled model spectra (thick line) at three different ages after the onset of 
a burst on an elliptical (E) galaxy with stellar population metallicity Z=0.008.
The best fits occurs for E($B-V$) in the range 0.36 - 0.41 mag and burst ages of 0.90 - 1.04 Gyr.
{\bf b)} SED for the Z=0.008 E model at various burst ages. Symbols as in Fig.~\ref{g2Eccsed}.} 
\label{Ez008}
\end{figure*}

To test the possibility that G2 is an early-type, gas poor galaxy, that experienced
an episode of star formation as a consequence of an interaction with 
or gas accretion from the companion(s),
we used a CC model with an E-type SFH and turned on an episode of star formation
at a galaxy age of 11 Gyr (the results would not change appreciably for an onset of the burst
at later ages). At such a late age the gas reservoir of the E-type model-galaxy has been 
replenished by evolved stars, and it is therefore possible to simulate relatively strong
bursts. We could not use an S0 model for this purpose, since the SFH assumed for S0 leaves
the galaxy with too small a reservoir of gas at late ages, so that a sufficiently strong burst 
to explain the observed SFR cannot be simulated. Since galaxies are treated as closed boxes, 
we could not simulate gas accretion from the companion galaxies.
The color evolution for E-type models is shown in Fig.~\ref{colevolE}.
We simulated burst strengths of 27\%, 15\%, and 9\% with e-folding timescale of 2$\times$10$^8$ yr.
An analogous fitting procedure to that described in Sect. 3 was applied and the best
value for internal extinction was found to be in the same range as for Sa and Sb models
(Sect. 4.1).
In all cases a good agreement between modeled and observed spectrum is reached 
at increasingly old burst ages for increasing burst strengths, in the range 1.3 to 1.6 Gyr 
(Fig.~\ref{g2Eccspec}), when the SFR of the model has dropped far below the observed value.
Model metallicities, $\approx$ 0.2 Z$_{\sun}$, are lower than observed and
model SEDs, contrary to the case of Sa or Sb models, are not in good agreement with 
NIR photometry of G2 (Fig.~\ref{g2Eccsed}). 
In fact a non-CC model with elliptical-like SFH, but with metallicity of the stellar populations set
to Z=0.008 (i.e. 0.4 Z$_{\sun}$), gives younger burst ages and a better agreement with the 
observed SED (Fig.~\ref{Ez008}). 
Such a non-CC model could be used to mimic the case in which the early-type galaxy
has been accreting some relatively high-metallicity gas from the companion(s), while
still keeping at a minimum the number of free parameters using a closed box model.
For this last model, with a burst strength of 11\%, the best 
agreement with the observed spectrum is found at burst ages 0.9 - 1.0 Gyr. 
Taking as reference the $K$ band luminosity and the model mass-to-light ratio,
we derive a total mass of 7$\times$10$^{10}$ M$_{\sun}$ and a stellar mass of 
4$\times$10$^{10}$ M$_{\sun}$ for G2 (i.e. about twice as high as derived from the Sa model). 
Scaling the model SFR accordingly
gives SFR $\sim$ 0.1 -  0.05 M$_{\sun}$ yr$^{-1}$ for 0.9 and 1.0 Gyr burst ages, respectively.
These values are lower than the measured 0.4 - 0.8 M$_{\sun}$ yr$^{-1}$. However, measured values
suffer from a high uncertainty in the extinction determination (because of the extreme
weakness of the H$\beta$ emission-line), therefore we still consider the model in reasonable agreement
with the observations. In Table~\ref{Ecol} we show the scaled model luminosities,
masses, and SFRs and the modeled colors for different burst ages.
Luminosities are in reasonable agreement (within the errors) with the observed ones.
At ages in the range 0.9 - 1.0 Gyr after the onset of the burst, the newly formed stars 
contribute 8\% of the stellar mass of the galaxy and $\sim$ 32\% of its visual light.

\begin{table*}
\caption[]{G2 -- Observed$^{\mathrm{a}}$ and Z=0.008 E Model Colors and Luminosities}
\label{Ecol}
$$
\begin{tabular}{lllllllll}
\hline
\hline
\noalign{\smallskip}
Object/Model & $B-V$ & $V-R$ & $V-K$ & $B-R$ & $B-H$ & $J-H$ & $H-K$ \\
             & M$_{B}$ & M$_{V}$ & M$_{R}$ & M$_{J}$ & M$_{H}$ & M$_{K}$ & SFR (M$_{\odot}$ yr${-1}$)\\
\noalign{\smallskip}
\hline
\noalign{\smallskip}
G2 & 0.88 & 0.46 & 2.52 & 1.34 & 3.24 & 0.59 & 0.16 \\
   & $-$19.82 & $-$20.70 & $-$21.16 & $-$22.47 & $-$23.06 & $-$23.22 & 0.4 - 0.8 \\
 & & & & & & &  \\
Z=0.008 E, $b$ = 0.11, 760 Myr 
   & 0.60 & 0.54 & 2.67 & 1.14 & 3.02 & 0.61 & 0.24 \\
M = 6.75$\times$10$^{10}$ M$_{\odot}$   & $-$19.96 & $-$20.55 & $-$21.09 & $-$22.37 & $-$22.98 & $-$23.22  &  0.20 \\
Z=0.008 E, $b$ = 0.11, 900 Myr    
   & 0.64 & 0.56 & 2.68 & 1.20 & 3.09 & 0.61 & 0.24 \\
M = 6.94$\times$10$^{10}$ M$_{\odot}$   & $-$19.90 & $-$20.54 & $-$21.10 & $-$22.38 & $-$22.99 & $-$23.22  &  0.10 \\
Z=0.008 E, $b$ = 0.11, 1.04 Gyr    
   & 0.68 & 0.57 & 2.70 & 1.25 & 3.14 & 0.61 & 0.24 \\
M = 7.03$\times$10$^{10}$ M$_{\odot}$   & $-$19.84  & $-$20.52 & $-$21.09 & $-$22.38 & $-$22.98 & $-$23.22 & 0.05 \\
Z=0.008 E, $b$ = 0.11, 1.18 Gyr    
   & 0.72 & 0.59 & 2.72 & 1.30 & 3.20 & 0.61 & 0.23 \\
M = 7.03$\times$10$^{10}$ M$_{\odot}$   & $-$19.79  & $-$20.50 & $-$21.09 & $-$22.38 & $-$22.99 & $-$23.22 & 0.03 \\
\noalign{\smallskip}
\hline
\end{tabular}
$$
\begin{list}{}{}
\item[$^{\mathrm{a}}$] Observed luminosities and colors after the application of a standard inclination correction.
\end{list}
\end{table*}

\section{Multiple bursts}

\begin{figure*}
\centering
\hbox{
\includegraphics[width=0.55\linewidth,height=6cm]{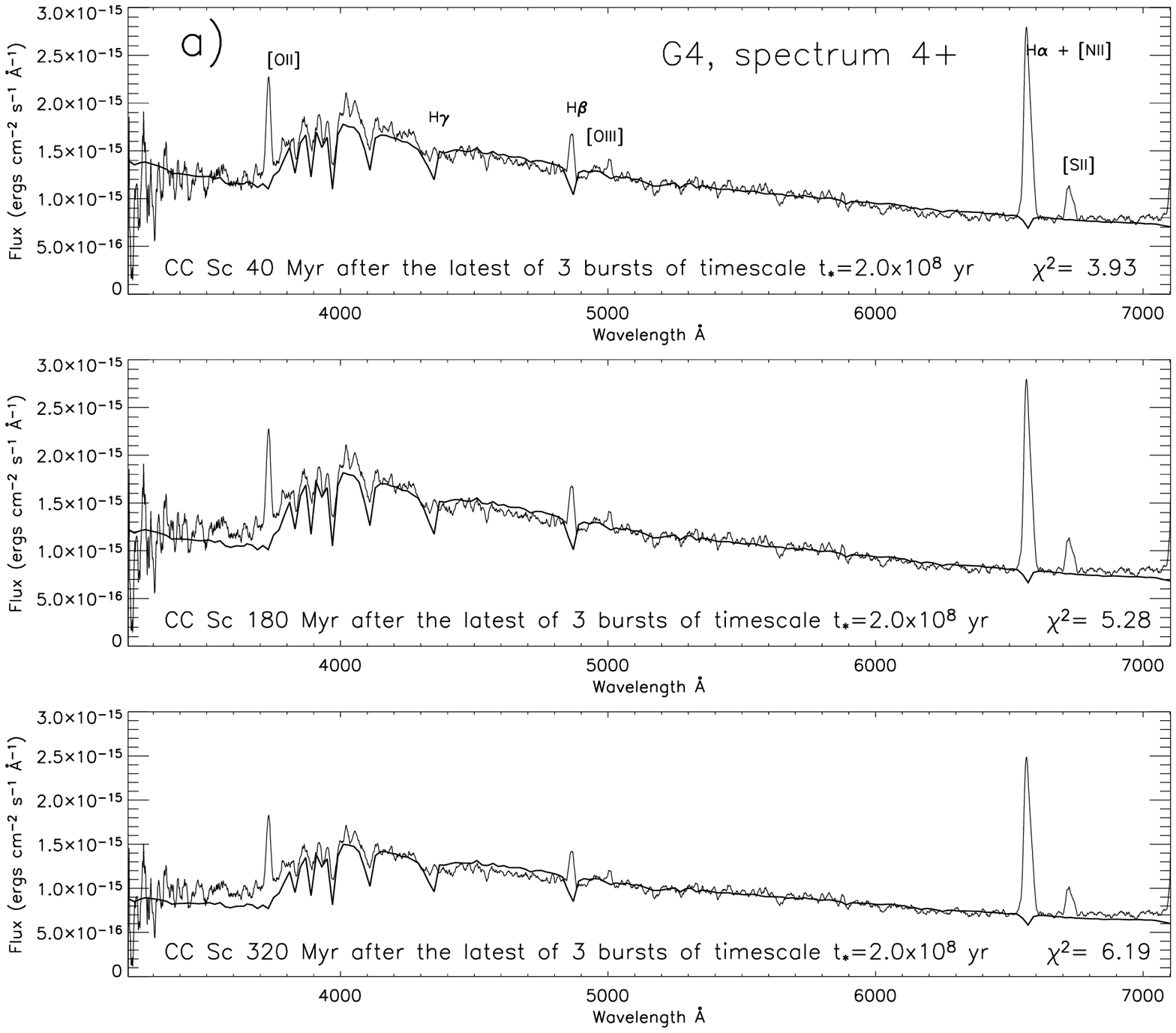} 
\includegraphics[width=0.45\linewidth,height=6cm]{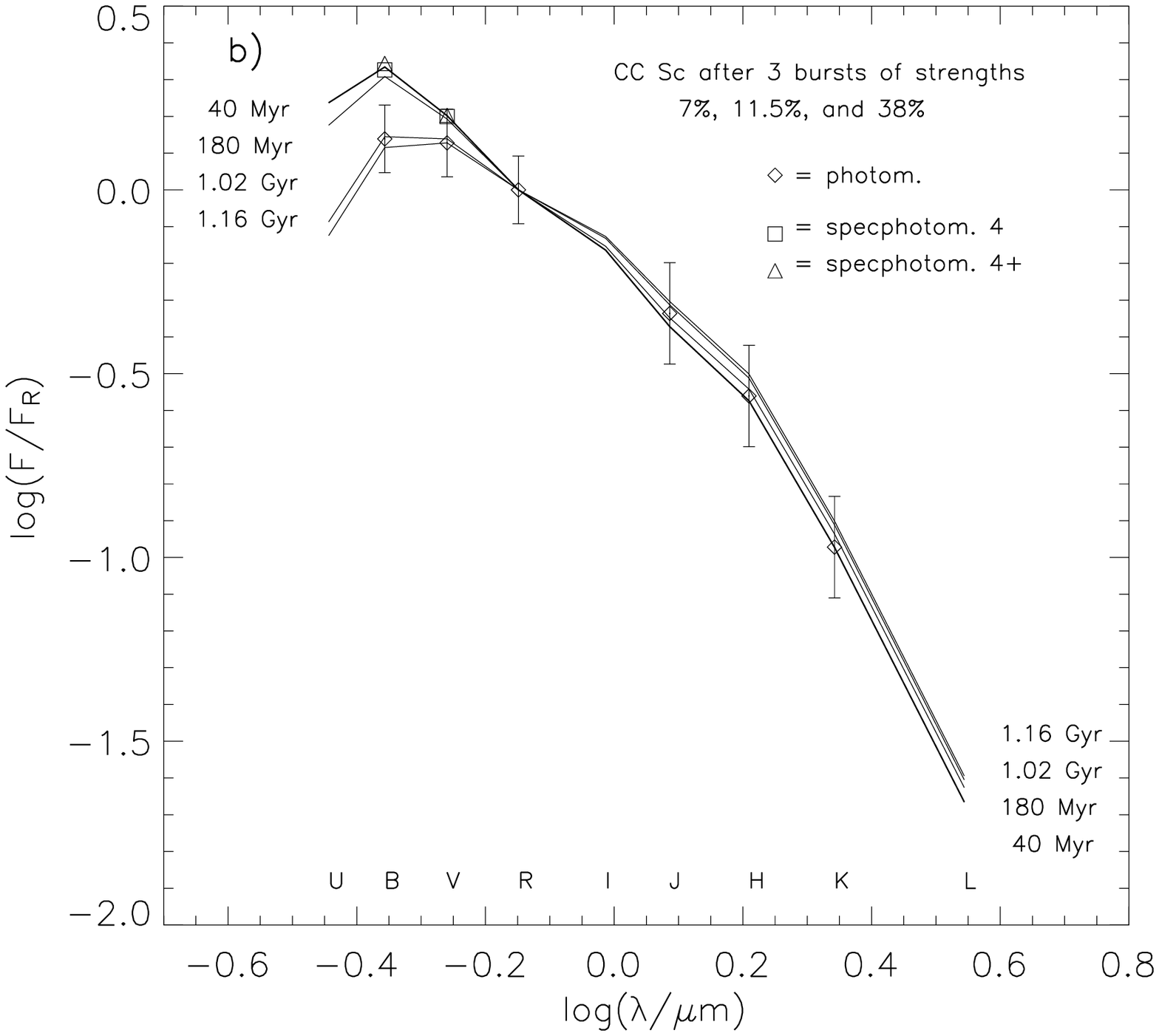}} 
\caption{{\bf a)} Spectrum of G4 at P.A. 90\degr\ corrected for internal reddening and overlapped with 
CC-Sc model spectra (thick line) at three different ages after the latest of three interaction-induced 
bursts of star formation (see text). 
{\bf b)} SED for the CC-Sc model at three ages after the latest of three interaction induced
bursts. Photometric points of G4 are marked with diamonds, while
the spectrophotometric points derived from the dereddened spectra at 
P.A. 130\degr\ and 90\degr\ are indicated by triangles and squares, respectively. The thick line
indicates the best fit age, according to the spectral analysis.} 
\label{triple-burst}
\end{figure*}

The evolutionary synthesis approach, as opposed to population synthesis, does not allow
to discover unexpected SFHs. In the models presented above, we have assumed the simplest
case of the occurrence of a single interaction-induced burst, which limits the
number of free parameters. However, in the tight galaxy system under study, the galaxies
may have undergone already several mutual interactions before the present one and, 
as a consequence, they might have experienced multiple bursts of star formation.  
We investigated the possibility to reproduce the observed spectrophotometric properties
of the two spiral galaxies assuming multiple bursts of star formation. 
However, it turned out that the observed spectra can be reproduced
only assuming a recent burst much stronger than the possible previous ones. Whatever the
age of previous interaction-induced bursts of star formation, satisfactory models are
obtained only 40 to 180 Myr after the latest strong burst. 
Also, simulating multiple bursts implies a considerably increase in the number of free parameters,
which cannot be constrained with the available observations. The consequent degeneracy in the models 
makes it impossible to establish with any confidence the relative age
and strength of possible previous bursts.
Therefore, our models do not
allow us to draw any conclusions on previous bursts. We can only conclude that the spectral
features are dominated by the latest strong burst.
As an example, we provide here the parameters of a triple-burst model whose match to the 
observational data of G4 would be as good as that of the single burst model presented in Sect. 3.
The model could be successfully applied also to G1 after rescaling to the appropriate total mass.
In this model the three bursts were switched on at a galaxy age of 11.0, 11.7, and 12.0 Gyr, respectively.
The burst duration was assumed to be the same, 2$\times$10$^8$ yr, for each burst. 
Burst strengths were set to 7\%, 11.5\%, and 38\%, respectively. The best match to the observed 
spectrum occurs 40 to 180 Myr after the latest burst, similarly to the single burst model
(reduced-$\chi^2$ = 3.93 and 5.28 for burst ages of 40 and 180 Myr, respectively, 
and best fit value of internal extinction E($B-V$) = 0.49 mag for the spectrum 4+).
For the model 40 Myr after the burst, the mass fractions of stars with ages t $<$ 40 Myr,
40 Myr $\leq$ t $\leq$ 340 Myr, and 440 Myr $\leq$ t $\leq$ 1 Gyr are
$\approx$ 7\%, 10\%, and 6\%, respectively, with the youngest stars contributing $\approx$ 30\% of the 
total $V$-band luminosity.
For the model 180 Myr after the latest burst, $\sim$ 20\% of stars (in mass) were produced after the
onset of the burst and they contribute $\sim$ 47\% of the light in the V band.
SFRs calculated 40 and 180 Myr after the latest burst are $\sim$ 12 and 6 M$_{\odot}$ yr$^{-1}$,
respectively, and the relevant metallicities are 0.25 and 0.36 Z$_{\odot}$.
Therefore, also in case of multiple burst models, all information taken together points to
an age of the latest burst between 40 and 180 Myr.
For models with previous bursts considerably stronger than the latest one, no agreement is found between 
observations and models at any age.
The comparison of multiple-burst model spectra and SEDs at different ages to the observed spectroscopic and
photometric data is shown in Fig.~\ref{triple-burst}. The difference with respect to 
the single-burst models shown in Figs.~\ref{g4+CC2spec}, \ref{g4CCsed} is only marginal. 
Colors, luminosities and metallicities of the 
triple-burst model 40 and 180 Myr after the latest burst are given in Table~\ref{comp3bG4}.
Differences between colors obtained with the single burst models and those obtained with the triple-burst
model are well within the errors. Taking as reference the model with older burst age, provided 
the modeled mass-to-light ratio is valid for G4, the observed luminosity would imply a stellar mass 
of 1.5$\times$10$^{10}$ M$_{\odot}$.
The accordingly scaled SFR of the model would be $\sim$ 5.5 M$_{\odot}$ yr$^{-1}$, still in 
reasonable agreement with the observations, while the metallicity reached by the model is higher 
than observed.
Slightly higher total and stellar masses ($\sim$ 2.4$\times$10$^{10}$ M$_{\odot}$ and
$\sim$ 1.8$\times$10$^{10}$ M$_{\odot}$, respectively) would be implied by the 40 Myr old
burst model, but the model SFR would be too high with respect to the observations ($\sim$ 14.9 
M$_{\odot}$ yr$^{-1}$).

\begin{table*}
\caption[]{G4 -- Observed$^{\mathrm{a}}$ and Triple-burst Model Colors and Luminosities}
\label{comp3bG4}
$$
\begin{tabular}{lllllllll}
\hline
\hline
\noalign{\smallskip}
Object/Model & $B-V$ & $V-R$ & $V-K$ & $B-R$ & $B-H$ & $J-H$ & $H-K$ \\
             & M$_{B}$ & M$_{V}$ & M$_{R}$ & M$_{J}$ & M$_{H}$ & M$_{K}$ &Z/Z$_{\odot}$ \\
\noalign{\smallskip}
\hline
\noalign{\smallskip}
G4 & 0.55 & 0.51 & 2.20 & 1.06 & 2.57 & 0.48 & 0.18 \\
   & $-$20.30 & $-$20.86 & $-$21.36 & $-$22.40 & $-$22.87 & $-$23.06 & $\sim$0.1 \\
 & & & & & & &  \\
CC-Sc, 3 bursts, 40 Myr 
   & 0.26 & 0.34 & 1.98 & 0.61 & 2.04 & 0.52 & 0.20 \\
M = 2.38$\times$10$^{10}$ M$_{\odot}$    & $-$20.81 & $-$21.08 & $-$21.41 & $-$22.34 & $-$22.85 & $-$23.06  & 0.25 \\
CC-Sc, 3 bursts, 180 Myr 
   & 0.25 & 0.32 & 2.01 & 0.57 & 2.05 & 0.53 & 0.22 \\
M = 1.77$\times$10$^{10}$ M$_{\odot}$    & $-$20.80 & $-$21.05 & $-$21.37 & $-$22.32 & $-$22.85 & $-$23.06  &  0.36 \\
\noalign{\smallskip}
\hline
\end{tabular}
$$
\begin{list}{}{}
\item[$^{\mathrm{a}}$] Observed luminosities and colors after the application of a standard inclination correction.
\end{list}
\end{table*}

\section{Discussion and conclusions}

We have used the G\"ottingen evolutionary synthesis code {\sc galev} to obtain 
chemically consistent spectrophotometric evolutionary synthesis
models to be compared to the three strongly interacting, star forming galaxies in
the compact group CG~J1720-67.8.

Our goal was finding the simplest solutions accounting for the observed
properties of the galaxies, within the observational and model uncertainties. 
The models we have presented and compared with 
the observational data were obtained under a number of simplifying assumptions,
like considering each galaxy as a closed box and assuming an undisturbed evolution
-- with a SFH consistent with the morphological type of the galaxies --
before the present/last encounter/starburst event. These assumptions are justified by the
need to keep the number of free parameters as small as possible, because of the
impossibility of constraining a larger number of parameters with the limited amount
of data at our disposal. As a result, the accuracy of the solutions presented here
is necessarily limited and some degree of degeneracy between the model parameters
(e.g. progenitor type and/or actual value of internal extinction and burst strength) still remains. 
The interaction history of CG1720-67.8 probably involves multiple encounters and
induced bursts of star formation. Therefore, we have investigated the possibility of occurrence
of multiple bursts for the spiral galaxies. However, no unique solution can be found
with presently available data, because of the high number of additional free parameters involved, 
like the duration of the individual bursts, their number and time separation and their relative
strengths. We provided an example of multiple burst model which would give as good a solution
as the simpler single burst model applied to G4. Although we cannot establish how many
interaction-induced bursts of star formation have occurred in the galaxy history, our models
indicate that the presently observed properties are reproduced at best when the latest 
burst episode is assumed to be stronger than previous bursts.
This result is consistent with the widely accepted belief that major bursts
destroy galaxy disks, which at best can be rebuilt in a timescale of 3 - 4 Gyr as
small, Sa-type disks. Since galaxies G1 and G4 both still show significant disks,
we can rule out the possibility that they have undergone particularly strong
bursts in the last 3 Gyr, before the presently observed one.
A set of data spanning a wider wavelength range, from UV to NIR, as well as 
a dynamical model of the entire galaxy group would provide constraints for additional parameters,
giving a basis for more sophisticated evolutionary synthesis models, which could include,
besides multiple bursts, mass transfer or exchange between galaxies or with the intragroup
medium. Work to build a dynamical model of the galaxy group
is presently in progress. A further limitation to our models is given by the lack of an 
accurate knowledge of the internal extinction, as discussed in Sects. 3 and 4.
However, with the above caveats in mind and taking into account the uncertainties (both
in the observations and the models), we were able to obtain models reasonably consistent
with the observational data. These give us an indication of the age and strength of the
(latest) interaction-induced star formation episodes in the galaxies and yield to 
estimates of the stellar masses of the galaxies.

In the version of {\sc galev} we used, the contribution of gaseous emission to broadband fluxes
was not included. However, the models presented here have burst ages $\geq$ 40 Myr.
At these ages the contribution of ionized gas emission (both in terms of lines and continuum)
to broadband fluxes is negligible at any metallicity, as shown by \citet{afva03}.

We found that the two late-type group's spirals, G1 and G4, have SEDs and optical spectra 
in reasonable agreement with a CC-model of an Sc galaxy at an age $\sim$ 40 -- 180 Myr after 
the onset of an interaction-induced intermediate to strong burst of star formation.
In particular the strong and older burst offers the preferred solution for G1, while
intermediate ages and strengths between the two proposed models might be more suitable
to G4. From the $K$-band luminosity of G4 and the mass-to-light ratio 
of the preferred model we derived a stellar mass of $\sim$ 1.6$\times$10$^{10}$ M$_{\sun}$ 
and a total (stellar + gaseous) mass of $\approx$ 2$\times$10$^{10}$ M$_{\sun}$.
This luminous mass estimate appears reasonable when compared with the dynamical mass
of 2.2$\times$10$^{10}$ M$_{\sun}$
derived in Paper~IV from the rotation curve of the galaxy within a radius of 5\arcsec,
nearly one disk scale length. Using the fitting law to the rotation curve and
extrapolating to a radius of 8\arcsec (i.e. nearly two disk scale lengths) would
give a dynamical mass of 3.3$\times$10$^{10}$ M$_{\sun}$.
For G1 we took as reference the $R$-band luminosity, because measured NIR luminosities 
are contaminated by an overlapping, foreground M-type star and $BV$ luminosities are
more affected by extinction. The estimated total mass of G1 is in the range 4 to 
6$\times$10$^{9}$ M$_{\sun}$ .
We estimated that the burst increased the stellar mass of these galaxies by 9 to 24 per cent.

For the early-type galaxy G2, we found two alternative solutions, which
lead to significantly different estimates of the galaxy's total mass.
The first model explores the possibility that the galaxy results from the merging 
of two gas-poor early-type spirals, an idea that was suggested in Paper~IV on the basis
of the observational properties of the group, including its tidal features.
Data appear consistent with a model in which two Sa galaxies, each one $\sim$ 1.7$\times$10$^{10}$  
M$_{\sun}$ in mass, experienced a strong -- as far as their gas reservoir allowed -- 
burst of star formation during their merging process, $\sim$ 0.7 - 0.9 Gyr ago, increasing 
the stellar mass of the galaxy by $\sim$ 5\%. However, two Sb progenitors of $\sim$ 1.4$\times$10$^{10}$
M$_{\sun}$ each would also offer an acceptable solution and would imply a burst age of
$\sim$ 1.2 - 1.3 Gyr and a stellar mass increase of $\sim$ 11\%.

The alternative model considers the possibility that G2 was born as an early-type galaxy and recently 
experienced a central burst of star formation because of gas inflow induced by the interaction with 
its companions. A solution reasonably consistent with the data was found using a Z = 0.44 Z$_{\sun}$
model with E-type SFH and indicates a burst age in the range 0.9 - 1.0 Gyr, with a contributed
stellar mass of 8\%. The total mass derived for G2 is of 3.4$\times$10$^{10}$ M$_{\sun}$ for the
CC-Sa merger model and 7$\times$10$^{10}$ M$_{\sun}$ for the non-CC E model.

The burst ages we found for the two spiral galaxies are largely consistent with our rough 
estimate of the age of the long tidal tail responsible for $\sim$ 30\% of the group's optical luminosity.
This supports the idea that the tail was issued as a consequence of a recent interaction between the
two spirals. Although it is not possible to establish whether the S0 galaxy was formed
through a late merger event or was born as an early-type galaxy, the models seem to indicate 
that the episode of star formation, whose traces are still present in the galaxy spectrum,
is relatively old with respect to the burst age of the two companion galaxies.
This suggests that G2 is not involved in the interaction which produced the long tidal tail, but
might be responsible for the fainter tidal features visible in the group that can be interpreted 
as fading remnants of older interaction processes.
Detailed dynamical models should further help disentangling the complex interaction history
of this galaxy group.
The total (stellar plus gaseous) mass of the galaxies involved is $\ga$ 6$\times$10$^{10}$ M$_{\sun}$,
without considering the material contained in the main tidal tail, whose contribution,
judging from its percentage luminosity, is non-negligible. Therefore, the visible mass of the final 
group remnant is expected to be of order of $\la$ 10$^{11}$ M$_{\sun}$.

In the following, we discuss our results in a more general context, by comparing them to
published works on interacting galaxies and/or small galaxy groups.
\citet{bgk03}, although following a different approach from the one we adopted in this work,
estimated the ages, strengths, and timescales of bursts of induced star formation at the center of galaxies
in pairs. They found indication that the bursts in the galaxies with small
separations and large H$\alpha$ equivalent widths are typically younger and stronger than in other galaxies.
They concluded that the strengths and ages of triggered starbursts
probably depend on galaxy separation on the sky and suggested that low-mass galaxies experience stronger
bursts of triggered star formation. Their observations appeared consistent with bursts of star formation
switched on at close galaxy-galaxy pass, which continue and age while the galaxies move apart.
The results we presented in this work for CG J1720-67.8 appear consistent with the scenario
depicted by \citet{bgk03}.

With our work we probed processes at work in a very compact galaxy configuration, found in isolation,
that we believe to be approaching coalescence.
It is unclear whether a similar behavior is to be expected for analogously compact configurations
embedded in denser environments. Recent investigations by \citet{coz04} suggest that in such cases
group evolution might proceed from low velocity dispersion groups with active star formation,
through intermediate velocity dispersion groups containing a large fraction of interacting/merging 
galaxies, to high velocity dispersion groups dominated by inactive elliptical galaxies.
To clear up this point, detailed studies similar to the one on CG J1720-67.8 that we presented 
in this series of papers, but targeting very compact galaxy systems embedded in environments with
differing densities, would be particularly useful.

In a recent study, \citet{bal04} analyzed the dependence of star formation activity on environmental 
density by considering group catalogs extracted from the 2dF Galaxy Redshift Survey \citep[2dFGRS, ][]{col01} 
and the
Sloan Digital Sky Survey \citep[SDSS, ][]{ab03}. They found that the fraction of star-forming 
galaxies depends strongly on local density and also on large-scale density, approximately independently
of group velocity dispersion, while the distribution of H$\alpha$ equivalent width in the star-forming
population does not depend on the environment. However, their analysis excludes the particular category/evolutionary
phase of groups, of which CG J1720-67.4 is an example. In fact their study is limited to groups with
velocity dispersion $>$ 200 km s$^{-1}$ and with at least 10 member galaxies. Additionally, extremely 
compact galaxy configurations, like the one considered in our work, cannot be selected in redshift space 
from the SDSS and 2dFGRS, because of observational constraints which impose a lower limit to the projected
angular separation between galaxies for which spectra can be obtained.
At lower velocity dispersions and small galaxy separations, interaction processes are more effective and induced 
star formation can become important. It is this situation that is expected to dominate the coalescence
phase in group evolution, i.e. the phase leading to the formation of ``fossil'' groups.
In fact, in our work we have shown that the star formation activity of the galaxies in the studied group
is strongly related to their recent interaction history. Moderate to strong bursts of star formation have
been switched on in two of the three members at an age compatible with the time of the latest strong interaction,
namely within the last 200 Myr (as indicated by the length of the tidal tail). An older ($\approx$ 1 Gyr), not 
yet fully extinguished, star formation
episode located in the early-type galaxy suggests an older interaction or even the first merger event 
in the history of the group. 
As already discussed in Paper IV, the global star formation activity of CG J1720-67.8 is more intense, by at least 
a factor of 4, than that displayed by several well studied merging galaxy pairs. 
As a further term of comparison, we can consider the triple galaxy system \object{AM 1238-362}, whose properties
have been studied in detail by \citet{temp03c}. This last system as well is characterized
by very low velocity dispersion, but its median projected galaxy separation ($\sim$ 35 kpc) is
larger than that of CG J1720-67.8 and its members are all spiral galaxies, one of which hosting a
Seyfert 2 nucleus. Although morphological distortions in this small group are weak, at least two of
its members show enhanced star formation activity which has been suggested to be the effect of
a recent close passage in their orbits. Several candidate intergalactic \ion{H}{ii} regions,
another possible consequence of the interaction, have been identified in the immediate vicinity of
these galaxies \citep{temp05b}. Star formation, although more intense in the central regions of
the galaxies, is spread all across their disks, in analogy to what we observed in the spiral members
of CG J1720-67.8. SFR surface densities derived for the more active member
of the triplet, Tol 1238-364, from H$\alpha$ line emission measurements are comparable to those derived for
G1 and G4 in Paper II. With respect to AM 1238-362, CG J1720-67.8 appears to represent a later
evolutionary phase, in which galaxies have come closer and are undergoing more violent 
interactions.

By studying a subsample of compact groups from \citet{hi82}'s catalog,
\citet{ipv99} found no global enhancement of SFR in compact group members with respect to
a sample of field galaxies. However, for a few groups they found a ratio of H$\alpha$ to $B$ luminosity
significantly higher than values measured in synthetic groups built from their field galaxy sample.
Most of these groups contain interacting galaxy pairs with evident tidal features.
One of these is the well studied group, HCG 31, which is believed to have started its merging phase and shows 
similaraties with CG J1720-67.8 concerning its peculiar morphology, dynamical properties, star formation 
activity, and presence of candidate tidal dwarf galaxies \citep{am04,ls04,mdo05b}.
The star formation history of HCG 31 has been recently investigated by \citet{ls04}, who confirmed
previous claims of very young ages of the latest bursts of star formation, mostly $<$ 10 Myr, 
for the members of this group.
Considering our burst age estimates for the galaxies of CG J1720-67.8 and for its tidal dwarf galaxy candidates
($<$ 10 Myr, Paper IV), we suggest that the group is observed at a similar (or slightly later) phase in 
the merging process as HCG 31.

If CG J1720-67.8 is to be considered as representative of a late pre-merging phase in poor groups,
its properties suggest that sufficiently gas-rich groups should undergo a particularly active star-forming
phase before final coalescence.

\begin{acknowledgements}
This work was supported by the Austrian Science Fund (FWF) under projects P15065 and P17772. 
S. T. gratefully acknowledges hospitality at the Universit\"atssternwarte G\"ottingen
and travel support from the German Astronomische Gesellschaft. We are grateful to the anonymous referee
for his/her useful and constructive comments which helped improving the presentation of our work. 
S. T. is grateful to J. Fritz and S. Ciroi for helpful suggestions and discussions.
\end{acknowledgements}


\end{document}